\documentclass[letter, 10pt]{IEEEtran}

\newcommand{\ignore}[1]{}
\usepackage{fancyhdr}
\usepackage[normalem]{ulem}
\usepackage[hyphens]{url}
\usepackage{hyperref}

\usepackage[top=1in, bottom=1in, left=0.75in, right=0.75in]{geometry}

\fancypagestyle{firstpage}{
  \fancyhf{}

  \pagenumbering{arabic}
}

\title{Validating Simplified Processor Models in Architectural Studies} 
\author{
    \IEEEauthorblockN{Sizhuo Zhang, Andrew Wright, Daniel Sanchez and Arvind}\\
    \IEEEauthorblockA{
        Massachusetts Institute of Technology\\
        \{szzhang, acwright\}@mit.edu, \{sanchez, arvind\}@csail.mit.edu
    }
}

\usepackage{verbatim}
\usepackage{graphicx}
\usepackage{amsfonts} 
\usepackage{amsmath} 
\usepackage{color}
\usepackage{enumitem}
\usepackage{url}

\usepackage{subfig}
\newcommand{\ie}{\emph{i.e.}}

\newcommand{\etc}{\emph{etc.}}

\newcommand{\OneIPC}{1-IPC}
\newcommand{\TTP}{TTP}
\newcommand{\WSU}{WSU}
\newcommand{\LIBDN}{LI-BDN}

\mathchardef\mhyphen="2D  
\newcommand{\MathHyphen}{\mhyphen}
\newcommand{\OneIPCMath}{1\MathHyphen{}IPC}

\setlist{leftmargin=*}

\begin{document}

\maketitle
\thispagestyle{firstpage}
\pagestyle{plain}


\begin{abstract}

Cycle-accurate software simulation of multicores with complex microarchitectures is often excruciatingly slow. 
People use simplified core models to gain simulation speed. 
However, a persistent question is to what extent the results derived from a simplified core model can be used to characterize the behavior of a real machine. 

We propose a new methodology of validating simplified simulation models, which focuses on the trends of metric values across benchmarks and architectures, instead of errors of absolute metric values.
To illustrate this methodology, we conduct a case study using an FPGA-accelerated cycle-accurate full system simulator. 
We evaluated three cache replacement polices on a 10-stage in-order core model, and then re-conducted all the experiments by substituting a \OneIPC{} core model for the 10-stage core model. 
We found that the \OneIPC{} core model generally produces qualitatively the same results as the accurate core model except for a few mismatches. 
We argue that most observed mismatches were either indistinguishable from experimental noise or corresponded to the cases where the policy differences even in the accurate model showed inconclusive results. 
We think it is fair to use simplified core models to study a feature once the influence of the simplification is understood. 
Additional studies on branch predictors and scaling properties of multithread benchmarks reinforce our argument.
However, the validation of a simplified model requires a detailed cycle-accurate model! 

\end{abstract}

\section{Introduction} \label{sec: intro}

When designing a multicore processor, there are many architectural decisions to make at various levels of the design.
These decisions include number of cores; last level cache (LLC) capacity, line size, associativity, replacement policy, and distribution; cache hierarchy, coherency, and interconnects; lower level cache capacity, line size, associativity, and replacement policy; TLB size and associativity; branch predictors and training; multiply and divide latency and throughput; FPU latency and throughput; and many more decisions depending on the specific optimizations performed in the processor pipeline such as superscalar, out-of-order execution, and register renaming.

With all these decisions, it is desirable to have a quick way to evaluate different architectures.
Unfortunately cycle-accurate simulation of large multicore processors is very time-consuming.
Where a processor may be able to finish a benchmark in a couple of seconds, a cycle accurate simulator may take 3 to 5 orders of magnitude longer, ranging from hours to days depending on the complexity of the microarchitecture.

It is common for designers to sacrifice cycle-accuracy to gain simulation speedup, and to do this, they use approximate simulation methods including trade-offs like simplified processor and memory models, truncated simulation, and sampling.
These approximate simulation methods allow for the exploration of many architectures, but they introduce quantitative errors in the reported results.
If these errors are larger than the reported gains from an architectural improvement, then it is possible that the improvement shown in simulation does not translate to the real design.
A deeper problem is that these errors are difficult to characterize and therefore difficult to bound quantitatively.

It would be desirable to have an upper bound on these quantitative errors to prove the effectiveness of an architectural improvement, but it would require a cycle-accurate model running alongside the approximate simulator to give these errors with enough certainty.

Instead of finding bounds on quantitative errors, we could just focus on the accuracy of qualitative results obtained by these simulators, such as which LLC replacement policy is better.
If a simulator is shown to make comparisons between architectures accurately, then it could be used for accurate exploration of a large design space, and a cycle-accurate simulator could be used to simulate the final selection of parameters to get the accurate quantitative effects of the selected processor additions.

One simplified processor model that introduces a large quantitative error is the \OneIPC{} core approximation.
This core approximation takes only one cycle to execute an instruction if there is no cache miss.
While this core is clearly not useful to simulate differences in pipeline optimizations such as out-of-order execution, superscalar, and register renaming (since these reduce to the same \OneIPC{} model)---it could be useful to simulate architectural features outside of the pipeline such as those mentioned in the first paragraph.

\begin{figure}[!htb]
\centering
\includegraphics[width=0.48\textwidth]{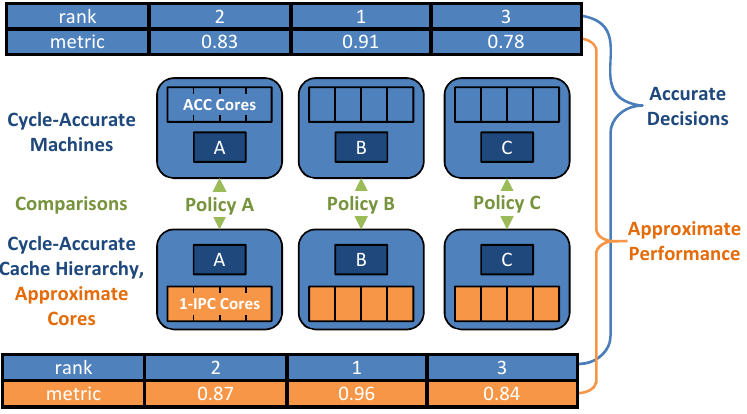}
\caption{Diagram of comparisons between accurate (ACC) core model and \OneIPC{} core model across three policies.} \label{fig: intro}
\end{figure}

This paper offers a careful study of the qualitative usefulness of a simplified \OneIPC{} core model by inserting it into a full system cycle-accurate simulator and comparing the results in various architectural studies to the results generated by the 10-stage in-order accurate core model (ACC).
Figure~\ref{fig: intro} shows a diagram of a typical experiment we would use to compare the two core models.
In this example, we are still able to accurately rank the policies by the inexact metric results from the approximate model.
Instead of comparing the simplified core model to a single reference machine, we are changing the machine with the simplified core model in tandem with the reference machine and show that the qualitative trends in each machine are very similar.

Our simplified \OneIPC{} core model is an approximation of a full system cycle-accurate PowerPC simulator: Arete \cite{khan2012fast}.
We chose Arete due to the cycle-accuracy and the ease of modification to create new machines to test new policies.
Since the \OneIPC{} processor is derived from a fully cycle-accurate simulator, the \OneIPC{} processor contains a cycle-accurate cache hierarchy and it runs the same OS as the cycle-accurate processor.
Since the core model is the only difference between the two systems, the effects of the \OneIPC{} core model approximation are isolated in our studies.

This paper makes a several key contributions to the study of approximate core models in architectural simulation:
\begin{enumerate}
	\item This paper proposes a new methodology of validating simplified simulation models, which focuses on the trends of metric values across benchmarks and architectures, instead of errors of absolute metric values.
	\item This paper presents an in depth, side-by-side comparison of a \OneIPC{} model with a cycle-accurate memory system against a fully cycle-accurate processor.
	\item This paper shows that two models agree for most cases, and in cases of mismatch, it is often when the magnitude of difference between the two choices is on the same order as the variations from run to run found in the cycle-accurate model.
	\item This paper employs a simple way previously proposed in the statistic field to estimate the variation of the cycle-accurate model.
\end{enumerate}

Throughout this paper we investigate the difference between the two models in various settings and across various metrics.
Section \ref{sec: relate work} covers related work.
Section \ref{sec: validate method} presents the our methodology to validate simplified models.
Section \ref{sec: simulator} includes an overview of the accurate and simplified core models and their implementation in the Arete simulator.
Section \ref{sec: llc} uses both models to evaluate three LLC replacement policies.
Section \ref{sec: scale} presents a study on the scalability of multithreaded benchmarks on processors varying from a single core up to 16 cores with both types of core models.
Section \ref{sec: brpred} compares three different branch predictors across the two core models for branch predictor accuracy.
Section \ref{sec: conclude} concludes the paper.

\section{Related Work} \label{sec: relate work}

Computer architects rely heavily on simulation to evaluate new techniques, and the community has developed numerous simulators with various degrees of accuracy. 
For example, GEMS \cite{martin2005multifacet}, M5 \cite{binkert2006m5} and MARSS \cite{patel2011marss} can achieve cycle-accuracy but run at relatively low speed, typically around hundreds of KIPS (thousand instructions per second).
Many other simulators trade off accuracy for higher speed, such as COTSon \cite{argollo2009cotson} which adopts a functional-directed simulation methodology.

In recent years, a class of simulators, such as CMP\$im \cite{jaleel2008cmp}, Graphite \cite{miller2010graphite}, Sniper \cite{carlson2011sniper}, ZSim \cite{sanchez2013zsim}, \etc, have been built on top of Pin \cite{luk2005pin}, a dynamic binary instrumentation framework developed by Intel.
Pin-based simulators are able to run at much higher simulation speed than sequential cycle-accurate simulators by leveraging native direct execution on the host machine, while still achieving good accuracy by using complex data structures and algorithms to track timing events. 
For example, Sniper is developed by combining the framework of Graphite with interval simulation \cite{genbrugge2010interval}, a technique that takes into account not only the delays caused by various miss events (such as cache miss and branch misprediction) but also the possible overlap of such miss events especially in out-of-order processors.
Though fairly accurate, these simulators are typically not cycle-accurate. 

For these non-cycle-accurate simulators, their designers have all carefully justified the simplifications that have been made, and may have also validated the simulator against the real machine. 
For example, ZSim has been validated against a 6-core Westmere machine.
However, the user of these simulators may simply take the simulator as accurate and will change the parameters and target architecture according to his needs.
It is possible that such changes will invalidate the designer's original justification about the simplifications in the simulator, because the user may be unaware of the designer's logic and argument.
Nowatzki \emph{et al.} \cite{nowatzki2014harmful} enumerate eight common pitfalls of some modern simulators, which may induce large simulation error if the user is unaware of them.
Therefore, whether a simulator with simplification can show similar trends as the real machine could be a problem.

There are several early studies focusing on simulator validation. 
Gibson \emph{et al.} \cite{gibson2000flash} compared applications' execution time derived from several FLASH simulators against the actual execution time on a real FLASH machine. 
Desikan \emph{et al.} \cite{desikan2001measuring} validated the sim-alpha simulator against a real Compaq DS-10L workstation, and they mainly focused on the IPC error. 
Cain \emph{et al.} \cite{cain2002precise} built a precise and accurate PowerPC processor model, and used it as the reference model in evaluating other simulators with simplifications. They conducted comparison using various metrics.

Our study differs from the previous work mainly in two aspects.
The first one is that we focus on whether the simplified model shows similar trends as the reference model while previous work concentrated on absolute errors in performance metrics. 
Although it is always useful to get accurate metric values from simulation, we argue that the primary goal of using a simulator is to explore new architectural changes and see whether they lead to improvement. 
Therefore, a simulator can be considered ``accurate" if it can qualitatively correctly predict the improvement caused by the change.
The second difference is that we apply architectural changes to our reference model (\ie{} the cycle-accurate model) since we want to see whether the simplified model can predict the effects of changes correctly. In previous studies no change was made to the reference model. 
The two studies by Gibson and Desikan were unable to do this since they used real machines which could not be modified as the reference model. 
Especially, Desikan \emph{et al.} did make changes to the target architecture but they could only evaluate the improvement on the simulators instead of the reference model (\ie{} the real machine).
If we want to evaluate the impact of an architectural change on the real machine, we will effectively need at least two machines, one with the change and other without.
Our cycle-accurate simulator has the flexibility that enables us to modify microarchitecture.
Although Cain \emph{et al.} also had such flexibility, they did not choose to do it.

Previous studies focusing on the inaccuracy of the \OneIPC{} core model have paid special attention to the influence of not modelling wrong path memory references in out-of-order cores. 
Mutlu \emph{et al.} \cite{mutlu2004understanding} studied the effects of wrong path memory references on the performance of an out-of-order superscalar uniprocessor. 
They argued that wrong path memory references can pollute the L2 cache and may act as prefetches. Ignoring them will cause a large error in the measured IPC. 
Sendag \emph{et al.} \cite{sendag2006quantifying} studied the impact of wrong path memory references on a 16-core (out-of-order) shared-memory multiprocessor. 
They showed a substantial portion of cache access, coherence traffic, replacement, \emph{etc.} are introduced by wrong path memory references.
Since our target architecture is an in-order core and all data memory accesses are non-speculative, we do not have these effects in our experiments.

Besides using simplified models, benchmark sampling is another technique to speed up architectural simulation.
Yi \emph{et al.} \cite{yi2005characterizing} evaluated two sampling techniques including SimPoint \cite{sherwood2002automatically} and SMARTS \cite{wunderlich2003smarts}, as well as other simulation techniques such as truncated simulation. 
They used a cycle-accurate simulator as baseline and applied two architectural enhancements to it separately. 
They then investigated whether they could see the same improvement when one of the simulation techniques was applied as the improvement they saw when none of them was used.
We conduct our experiments in a similar way.
In order to justify the usefulness of the simplified model, we will apply changes to the baseline architecture and see whether the simplified model can show the similar trends as the cycle-accurate model does.

\section{Methodology on Validating Simplified Models} \label{sec: validate method}

The absolute metric values measured on a simplified model generally cannot match those on the accurate model.
Instead of studying the error of absolute metric values, our validation methodology focuses on whether the simplified model matches the accurate model in terms of the relative trends of results across different benchmarks and architectures.

\noindent\textbf{Trends across benchmarks:}
When studying the trend of metric values across benchmarks, we fix the architecture in evaluation.
For each simulation model (\ie{} simplified and accurate), we construct a vector that contains the measured metric values of all benchmarks.
After normalizing the vectors, we can compare the distributions across workloads between the simplified model and the corresponding accurate model.
This shows how much the simplified model distorts the characteristics of the results across benchmarks.

\noindent\textbf{Trends across architectures:}
The primary goal of simulation is to compare two architectures and find out the better one.
For each simulation model and benchmark, we can compute the ratio of the metric values of two architectures. 
We refer to such ratios as \emph{improvement ratios}, because they represent the improvement of one architecture over the other.
We can now study the following questions about the fidelity of the simplified model:
\begin{enumerate}
	\item {Qualitatively, does the simplified model always agree with the accurate model in terms of which architecture is better for each benchmark?}
	In case of disagreement, we study the variation of the results of the accurate model.
	If the variation is so large that one may reach a wrong qualitative conclusion when he/she runs the experiments only once with the accurate model, then the architecture itself should be considered as \emph{brittle}.
    
	\item {Quantitatively, how much is the error of the improvement ratios from the simplified model compared to the run-to-run variation on the accurate model?}
	If the error is comparable to or even smaller than the variation, such error should not be viewed as problematic when we cannot run accurate simulation multiple times to reduce the variation.
\end{enumerate}

We will follow the above methodology to validate the simplified model using \OneIPC{} core in the case study of LLC replacement policy in Section \ref{sec: llc}.

\section{Simulator} \label{sec: simulator}

Our \OneIPC{} simulator and our accurate simulator are based off of the FPGA-based Arete simulator \cite{khan2012fast} (we obtained and modified the source code of Arete with the permission of the authors).
Arete is a full system simulator that implements the Power ISA--Embedded Environment and boots Linux.

The processor cores in Arete are 10-stage in-order pipelines modeled cycle-accurately using the Latency Insensitive Bounded Network (LI-BDN) technique \cite{vijayaraghavan2009bounded}. 
LI-BDN allows for refinements of the simulator implementation to reduce the FPGA resource budget and increase the clock speed while preserving cycle-accuracy.

\begin{figure}[htbp]
\centering
\includegraphics[width=0.4\textwidth]{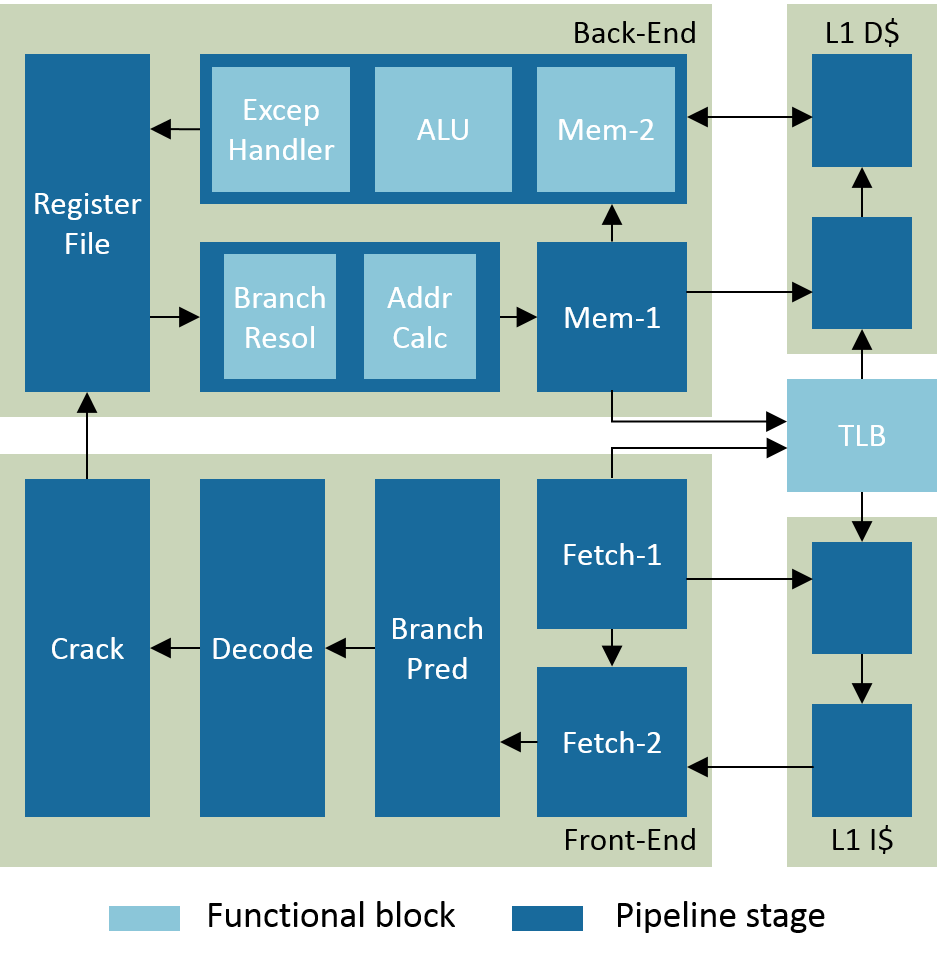}
\caption{Structure of 10-stage in-order pipeline} \label{fig: 10 stage pipeline}
\end{figure}

Figure \ref{fig: 10 stage pipeline} (copied from \cite{khan2012fast} with the permission of the authors) shows the structure of the 10-stage in-order pipelines.
The front end of the pipeline is five-stage long and includes instruction fetching, branch prediction, instruction decoding, and cracking complex instructions into multiple simple instructions.
The back end of the pipeline is also five-stage long and includes reading the register file, resolving branches, accessing memory, executing instructions, handling exceptions, and writing results back to the register file.
This pipeline does not include a floating point unit, so all benchmarks are compiled with software floating point operations. 

The original Arete simulator connects the core models through a shared L2 cache that lacks a detailed timing model.
For these experiments, we wanted cycle-accuracy at the processor-level, not just the core-level.
To get this, we expanded the LI-BDNs of the cores to include the L2 cache and main memory.
This resulted in detailed timing from the cores up to the memory hierarchy.

Table~\ref{tab: base-setup} shows the base simulator settings of a 4-core system and the timing models used for these experiments.
Each simulator used in each of the experiments was implemented on a VC707 FPGA board.

\begin{table}[!htb]
\centering
\begin{footnotesize}
\begin{tabular}{|l|l|}
\hline
Core & 4$\times$10-stage in-order pipeline, 2GHz frequency \\ 
& 256-entry branch target buffer (BTB) \\
& Tournament branch predictor from Alpha 21264 \cite{kessler1998alpha} \\ 
& 64-entry return address stack (RAS) \\
\hline 
L1 I cache & 4$\times$32KB, 4-way set associative, 64-byte block \\
& 1-cycle pipelined hit latency \\ 
& Blocking access (only 1 request in flight) \\
& True LRU replacement \\
\hline
L1 D cache & 4$\times$32KB, 4-way set associative, 64-byte block \\
& 1-cycle pipelined hit latency \\
& Blocking access (only 1 request in flight) \\
& True LRU replacement \\
\hline
L2 cache & 1$\times$2MB, 8-way set associative, 64-byte block \\
& Shared by all L1 caches, MSI coherence protocol \\
& 1-cycle tag access, 8-cycle pipelined data access \\
& At most 8 requests in flight \\
& True LRU replacement \\
\hline
Memory & 1$\times$2GB, 120-cycle access latency \\ 
& At most 12 requests in flight \\
& 12.8GB/s peak bandwidth \\
\hline
\end{tabular}
\end{footnotesize}
\caption{Base simulator settings of a 4-core system} \label{tab: base-setup}
\end{table}

The \OneIPC{} core model used throughout this paper only stalls during L1 I/D cache misses.
Otherwise it will issue, execute and commit each instruction in a single cycle.
This simplified core model is derived from the cycle-accurate core model through two steps.
First, all the FIFOs that connect adjacent pipeline stages are replaced with bypass FIFOs so that an instruction can flow through all stages in one cycle if there is no cache miss.
Second, appropriate stall logic is added that feeds into Fetch-1 and Crack stages in order to ensure that these two stages do not issue new instructions when there are still outstanding instructions in later pipeline stages.
With this stall logic, the whole pipeline will only have at most one in-flight instruction when the Crack stage is not active.
When the Crack stage is active, the stall logic ensures there is at most one outstanding instruction in the back end of the pipeline and one (\ie{} the complex instruction being cracked) in the front end.
By applying these 2 changes, the behavior of the pipeline will match our definition of the \OneIPC{} core model.

For the rest of the paper, we use ACC for the cycle-accurate 10-stage in-order core model for convenience, and we also use ACC model and \OneIPC{} model to denote the full system simulators that include ACC core models and \OneIPC{} core models respectively.
Note that the only difference between ACC model and \OneIPC{} model is the core model; the memory hierarchy and rest of the system is always simulated with cycle-accuracy.

\section{Study 1: Last Level Cache Replacement Policy} \label{sec: llc}

In this section, we evaluate three LLC replacement policies on the ACC and \OneIPC{} models.
For this experiment, we used the 4-core configuration shown in Table~\ref{tab: base-setup}.
We created three versions of this simulator, one for each policy.

\subsection{Candidate Policies}
This LLC replacement experiment compared the following three replacement policies.
\begin{enumerate}
\item \textbf{True LRU}.

\item \textbf{TADIP} (Thread Aware Dynamic Insertion Policy) \cite{jaleel2008adaptive}\cite{qureshi2007adaptive}. \\
This policy always evicts the LRU cache line, but inserts a new line into either LRU or MRU position.
It uses set duelling to dynamically select between two insertion policies: always inserting to MRU position and Bimodal Insertion Policy (BIP).
BIP inserts the new line into LRU position with probability $1-\epsilon$ and into MRU position with probability $\epsilon$.
We set $\epsilon$ to $1/32$, and we assign a MRU bit to each line to approximate the LRU replacement list.

\item \textbf{TADRRIP} (Thread Aware Dynamic Re-Reference Interval Prediction) \cite{jaleel2010high}. \\
This policy builds the replacement list using the 2-bit re-reference interval prediction value (RRPV) of each cache line.
It also use set dueling to dynamically select between two policies: a static one and a bimodal one.
The major difference between them is that the static one always initializes the RRPV of the new cache line to 2, while the bimodal one initializes it to 3 with probability $1-\epsilon$ and to 2 with probability $\epsilon$.
Here we also set $\epsilon$ to $1/32$.
\end{enumerate}
We implemented the set duelling mechanism with feedback described in \cite{jaleel2008adaptive} for both TADIP and TADRRIP. 
The important parameters are listed here:
\begin{itemize}
\item four 10-bit policy select counters, one for each core
\item 8 set duelling monitors (SDMs), two for each core
\item 32 dedicated cache sets for each SDM
\end{itemize}

For convenience, we use LRU, DIP and DRRIP to stand for true LRU, TADIP and TADRRIP in the rest of the paper.

\subsection{Benchmark}
We choose 6 single-thread applications from 3 benchmark suites as shown in Table~\ref{tab: benchmark}.
Four applications are taken from the SPEC CINT2006 benchmark suite, \texttt{pointer} is taken from the DIS Stressmark benchmark  suite \cite{DIS_stressmark}, and \texttt{stream} is taken from the STREAM benchmark suite \cite{mccalpin1995survey}.
\texttt{pointer} performs random memory accesses inside a 4 MB buffer. 
\texttt{stream} iterates through three arrays, each 781 KB long.

We only choose integer benchmarks here because floating operations are done in software due to the lack of floating point unit in Arete, so the portion of memory access instructions will become very low if we run floating point benchmarks.
We do not use other benchmarks in SPEC CINT2006 due to various reasons. Some benchmarks fail to stress the cache, some cannot be cross-compiled to PowerPC, and others take too long to finish.

\begin{table}[!htb]
	\centering
	\begin{footnotesize}
		\begin{tabular}{|c|c|p{0.2\textwidth}|}
			\hline
			Benchmark suite & Application & Input size and parameter \\
			\hline
			SPEC CINT2006 & bzip2 & train input, byoudoin.jpg \\
			\cline{2-3}
			& gobmk & test input, connect.tst \\
			\cline{2-3}
			& libquantum & train input \\
			\cline{2-3}
			& mcf & test input \\
			\hline
			DIS Stressmark & pointer & p12.in (in the new input set) \\
			\hline
			STREAM & stream & array element type \texttt{unsigned int}, array size 200000, kernel is repeated for 1000 times \\
			\hline
		\end{tabular}
	\end{footnotesize}
	\caption{Single-thread applications from 3 benchmark suites} \label{tab: benchmark}
\end{table}

With these 6 applications, we generate all possible 15 multiprogrammed application mixes for our 4-core system, according to the alphabetical order of single-thread application names.
Table \ref{tab: multiprog workload} shows part of the 15 multiprogrammed workloads.
We will evaluate the replacement policies using these 15 workloads.

\begin{table}[!htbp]
\centering
\footnotesize
\begin{tabular}{|c|l|l|l|l|}
\hline
Workload ID & Process 0 & Process 1 & Process 2 & Process 3 \\
\hline
1 & bzip2 & gobmk & libquantum & mcf \\
\hline
2 & bzip2 & gobmk & libquantum & pointer \\
\hline
\multicolumn{5}{|c|}{\ldots} \\
\hline
14 & gobmk & mcf & pointer & stream \\
\hline
15 & libquantum & mcf & pointer & stream \\
\hline
\end{tabular}
\caption{Multiprogrammed workload} \label{tab: multiprog workload}
\end{table}

\subsection{Measurement Methodology}
To test the LLC replacement policies, the multiprogrammed workloads listed in Table \ref{tab: multiprog workload} are run on each simulator where each process is pinned to a specific core.
For each single-thread application, two special instructions \texttt{progBeginTrap} and \texttt{progEnd} are inserted into the beginning and the end of the program respectively. 
When the process reaches the \texttt{progBeginTrap} instruction, the core will spin on this instruction until all the processes have reached this instruction. 
Simulation will terminate when all cores have executed the \texttt{progEnd} instruction at least once; programs that reach the \texttt{progEnd} instruction early restart. 
The number of cycles simulated for each workload ranges from 30 billion to 70 billion. 

We collect statistics from the moment when all cores leave the \texttt{progBeginTrap} instructions to the time that simulation terminates.
The following three metrics are measured for evaluation:
\begin{itemize}
\item L2 cache misses per 1000 instructions (MPKI);
\item total throughput $\mathrm{\TTP}=\sum_{i=0}^{3}\mathrm{IPC}_i$, where $\mathrm{IPC}_i$ is the instruction per cycle metric (IPC) for core $i$; and
\item weighted speedup $\mathrm{\WSU} = \sum_{i=0}^{3}\mathrm{IPC}_i / \mathrm{IPC_i^{single}}$, where $\mathrm{IPC}_i^{single}$ is the IPC metric for the program on core $i$ when the program runs in isolation on a single core with 512KB L2 cache using LRU policy.
\end{itemize}
In the rest of the paper, we will use MPKI, \TTP, and \WSU{} to refer to these these metrics in cases without ambiguity.

\subsection{Results}

In order to derive the WSU metric, we also ran all the single-thread programs seven times on both the ACC and \OneIPC{} models for the single-core processor with 512KB L2 cache using LRU policy.
Note that the WSU metric for the \OneIPC{} model should be calculated using the single-core IPC of the \OneIPC{} model.
We found the variation among the single-core IPC results is very small, because the standard deviation is less than 0.24\% of the mean value for each program.
Therefore we only use the mean values of the measured single-core IPC in the calculation of WSU, and ignore the variation of the single-core IPC.

\begin{figure}[!htb]
	\centering
	\subfloat[MPKI metric]{\includegraphics[width=0.45\textwidth]{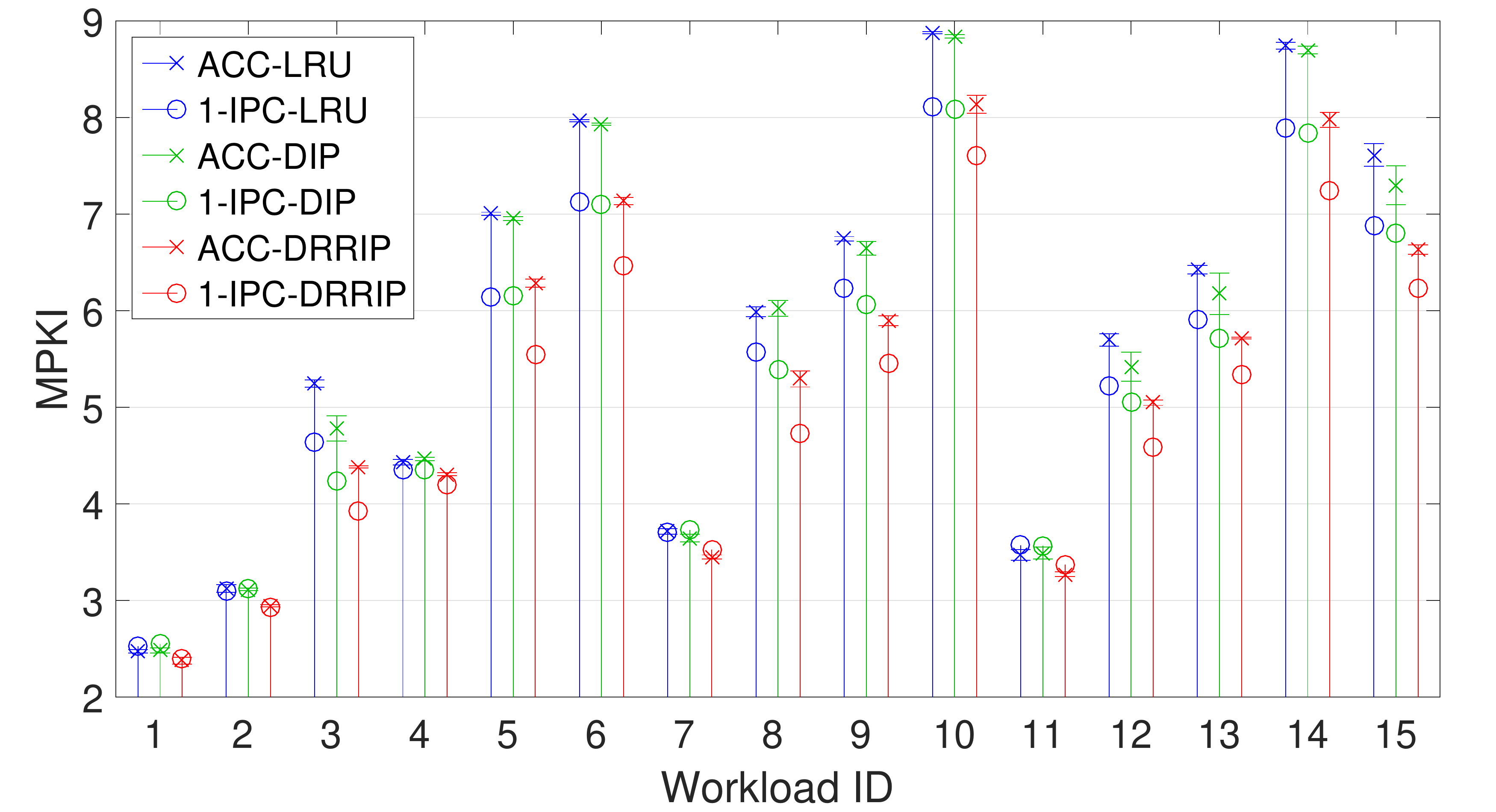}} \\
	\subfloat[\TTP{} metric]{\includegraphics[width=0.45\textwidth]{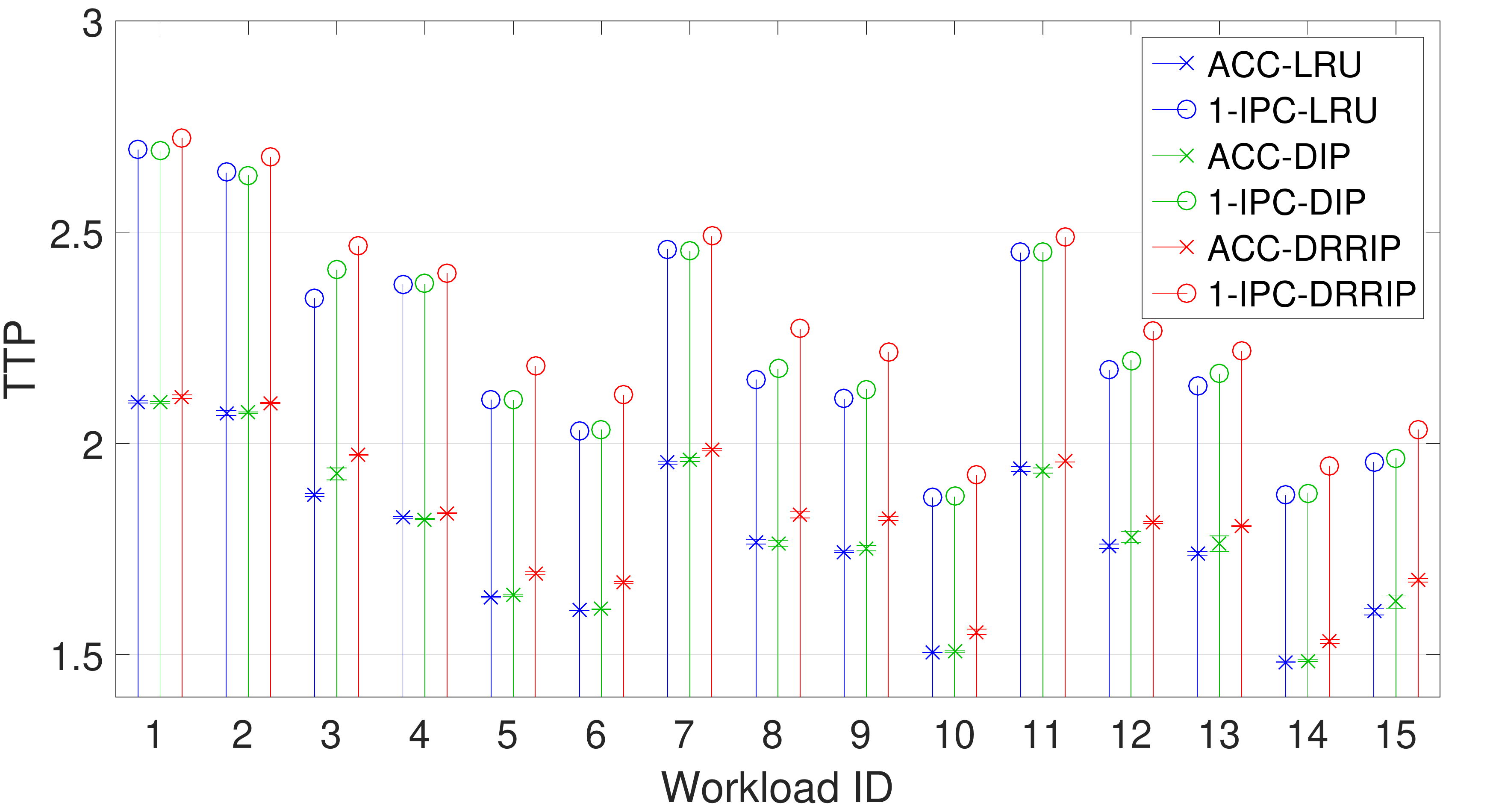}} \\
	\subfloat[\WSU{} metric]{\includegraphics[width=0.45\textwidth]{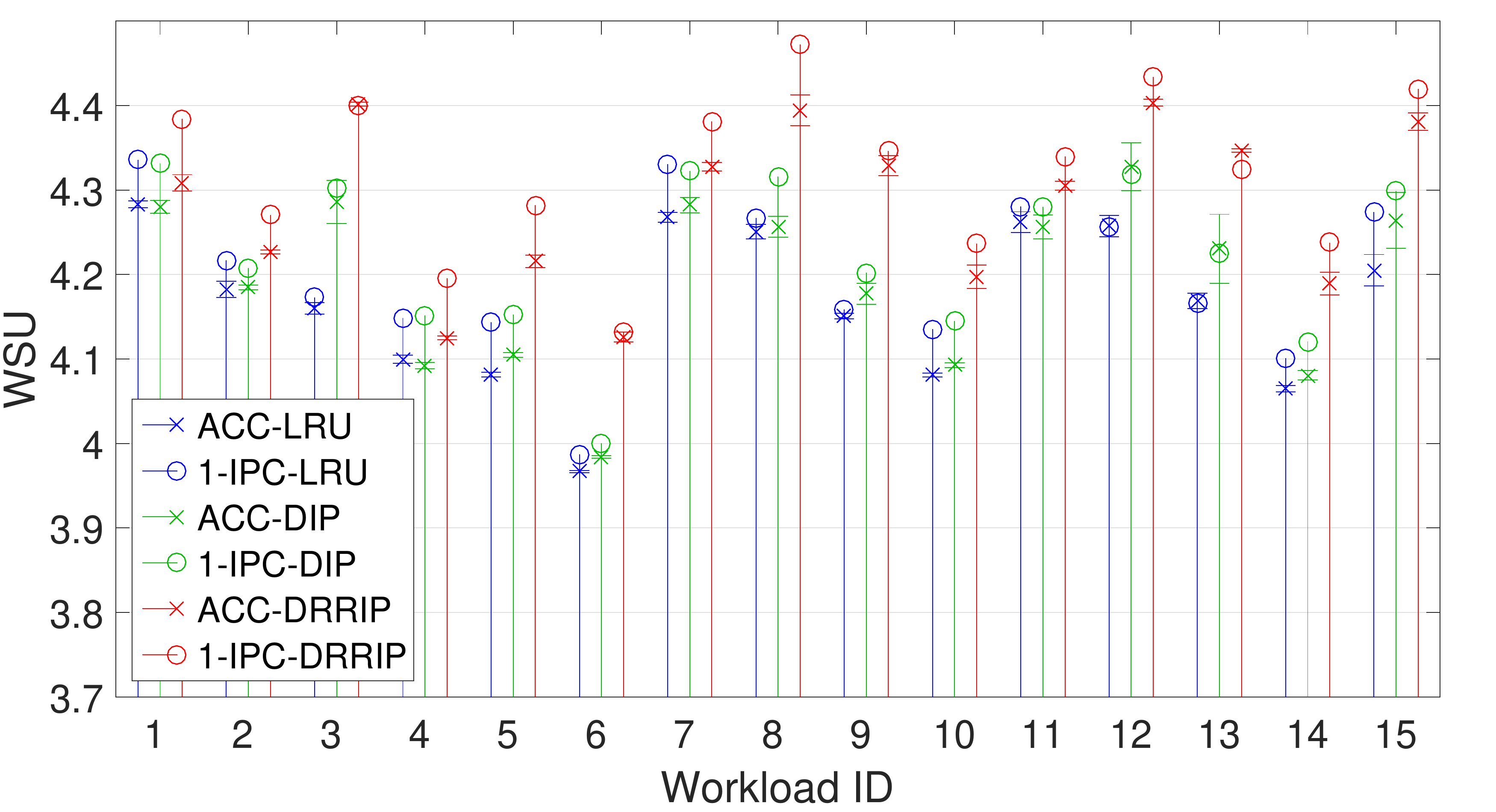}} 
	\caption{Mean values of three metrics over seven runs for each workload using each $\langle$model, policy$\rangle$ pair} \label{fig: abs mean result}
\end{figure}

As for the multiprogrammed workloads, we ran each of them for seven times on both the ACC and \OneIPC{} models. 
In particular, the multiple runs on the ACC model will capture the variability of the metrics mostly due to operating system effects and non-determinism in the replacement policies.
Figure~\ref{fig: abs mean result} shows the mean values (the cross markers) of the three metrics over all runs for each workload using each $\langle$model, policy$\rangle$ pair.
The standard deviations of the measured ACC metric values are represented by the distances from horizontal bars to cross markers, which are fairly small.

\subsection{Analysis}
In Figure \ref{fig: abs mean result}, we see that the results for the \OneIPC{} and ACC models are close for \WSU{}, while they fail to match in magnitude for MPKI and \TTP{}. 
Despite the difference in magnitudes, we will follow the methodology in Section \ref{sec: validate method} to show that two models exhibit the same trends across workloads and replacement policies.

\noindent\textbf{Comparing trends across workloads:}
We first focus on the difference between ACC and \OneIPC{} models for a fixed metric and policy.
To quantify the trend of metric $X \in \{\mathrm{MPKI},\mathrm{\TTP},\mathrm{\WSU}\}$ across workloads with the set policy, we form the vector $X_m \in \mathbb{R}^{15}$ containing the average metric values for each workload over all runs using model $m\in\{\mathrm{ACC}, \mathrm{\OneIPCMath}\}$.
To isolate the trends across workloads, we first normalize each $X_m$ vector so that its norm becomes 1. 
We then calculate the Euclidean distance between the normalized vectors of the two models (\ie{} $X_\mathrm{ACC}$ and $X_\mathrm{\OneIPCMath}$) to get an estimate of the similarity of their trends.

Table~\ref{tab: dis norm vec} shows the distances between the normalized metric vectors of two models for all combinations of policies and metrics. 
These distances are much smaller than 1, so using the \OneIPC{} model does not significantly distort the characteristics of the results for each workload.

\begin{table}[htbp]
\centering
\footnotesize
\begin{tabular}{|c|c|c|c|}
\hline
& LRU & DIP & DRRIP \\
\hline
MPKI & 0.0358  &  0.0375  &  0.0369 \\
\hline
\TTP & 0.0214  &  0.0207  &  0.0212 \\
\hline
\WSU & 0.0056  &  0.0050  &  0.0066 \\
\hline
\end{tabular}
\caption{Distance between normalized metric vectors of ACC and \OneIPC{} models for each policy and metric} \label{tab: dis norm vec}
\end{table}

\noindent\textbf{Comparing improvement ratios:}
Next we explore the difference between using ACC and \OneIPC{} models when comparing two policies.
The improvement ratio is the ratio of the metric value of the new policy over that of the baseline policy.
Besides an average ratio calculated using the mean values of the metrics shown in Figure \ref{fig: abs mean result}, we can estimate the variation of the improvement ratios on the ACC model, which will be later on compared with the error induced by the inaccuracy of the \OneIPC{} model.

Assume $x$ and $y$ are random variables that represent the metric values of the new policy and the baseline policy respectively for the same workload on the ACC model.
Then the improvement ratio will be $x/y$.
We further assume $x$ and $y$ follow normal distributions $N(\mu(x), \sigma(x))$ and $N(\mu(y), \sigma(y))$ respectively. 
The accurate probability distribution of $x/y$ is very complex, but we can simplify it based on the observation that the variations of metric values are much smaller than the absolute metric values as shown in Figure \ref{fig: abs mean result}.
Therefore we can perform a Taylor expansion on $x/y$ as follow ($\delta(x)$ and $\delta(y)$ represent $x-\mu(x)$ and $y-\mu(y)$ respectively) \cite{franz2007ratios}:
\begin{displaymath}
\frac{x}{y} = \frac{\mu(x)+\delta(x)}{\mu(y)+\delta(y)} \approx \frac{\mu(x)}{\mu(y)} \left( 1 + \frac{\delta(x)}{\mu(x)} - \frac{\delta(y)}{\mu(y)} \right) 
\end{displaymath}
Then $x/y$ approximately follows a normal distribution, in which $\mu(x/y) = \mu(x)/\mu(y)$ and
\begin{displaymath}
\sigma\left(\frac{x}{y}\right) = \frac{\mu(x)}{\mu(y)} \sqrt{\left( \frac{\sigma(x)}{\mu(x)} \right)^2 + \left( \frac{\sigma(y)}{\mu(y)} \right)^2}.
\end{displaymath}
We can estimate $\mu(x/y)$ and $\sigma(x/y)$ by substituting $\mu(x)$, $\mu(y)$, $\sigma(x)$ and $\sigma(y)$ with the mean values and standard deviations of measured ACC results shown in Figure \ref{fig: abs mean result}.

We use range $\mu(x/y)\pm1.28\sigma(x/y)$ as an estimate of the run-to-run variation of the ACC improvement ratio.
Since the probability of falling into this range is 80\% for the normal distribution of $x/y$, if one conducts the experiment only once, there will be 10\% possibility for the result to be larger than the whole range, and another 10\% possibility for the result to be smaller than the whole range.

\begin{figure}[!htb]
	\centering
	\subfloat[Using MPKI for comparison \label{fig: mpki dip/lru}]{\includegraphics[width=0.45\textwidth]{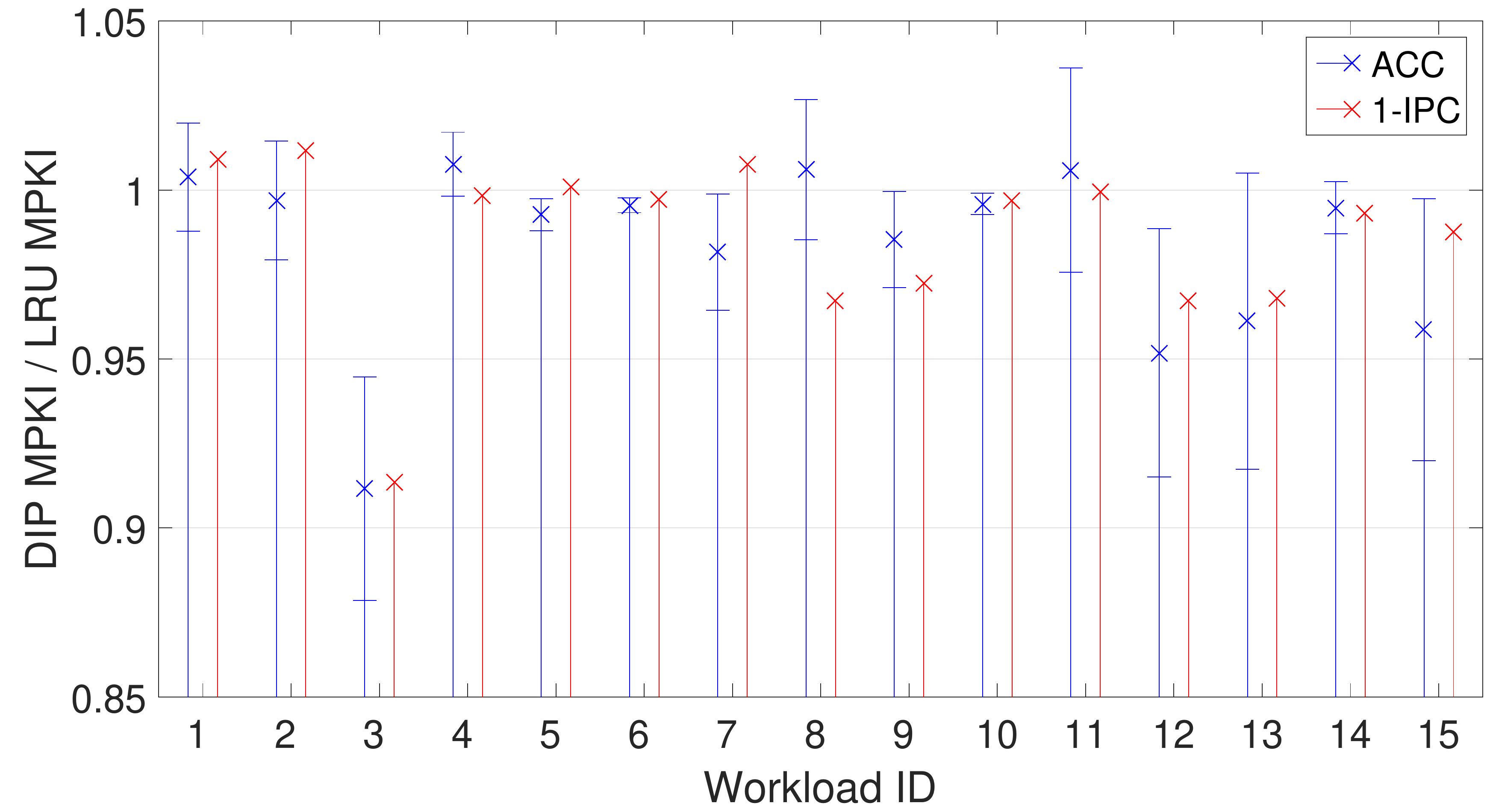}} \\
	\subfloat[Using \TTP{} for comparison \label{fig: ipc dip/lru}]{\includegraphics[width=0.45\textwidth]{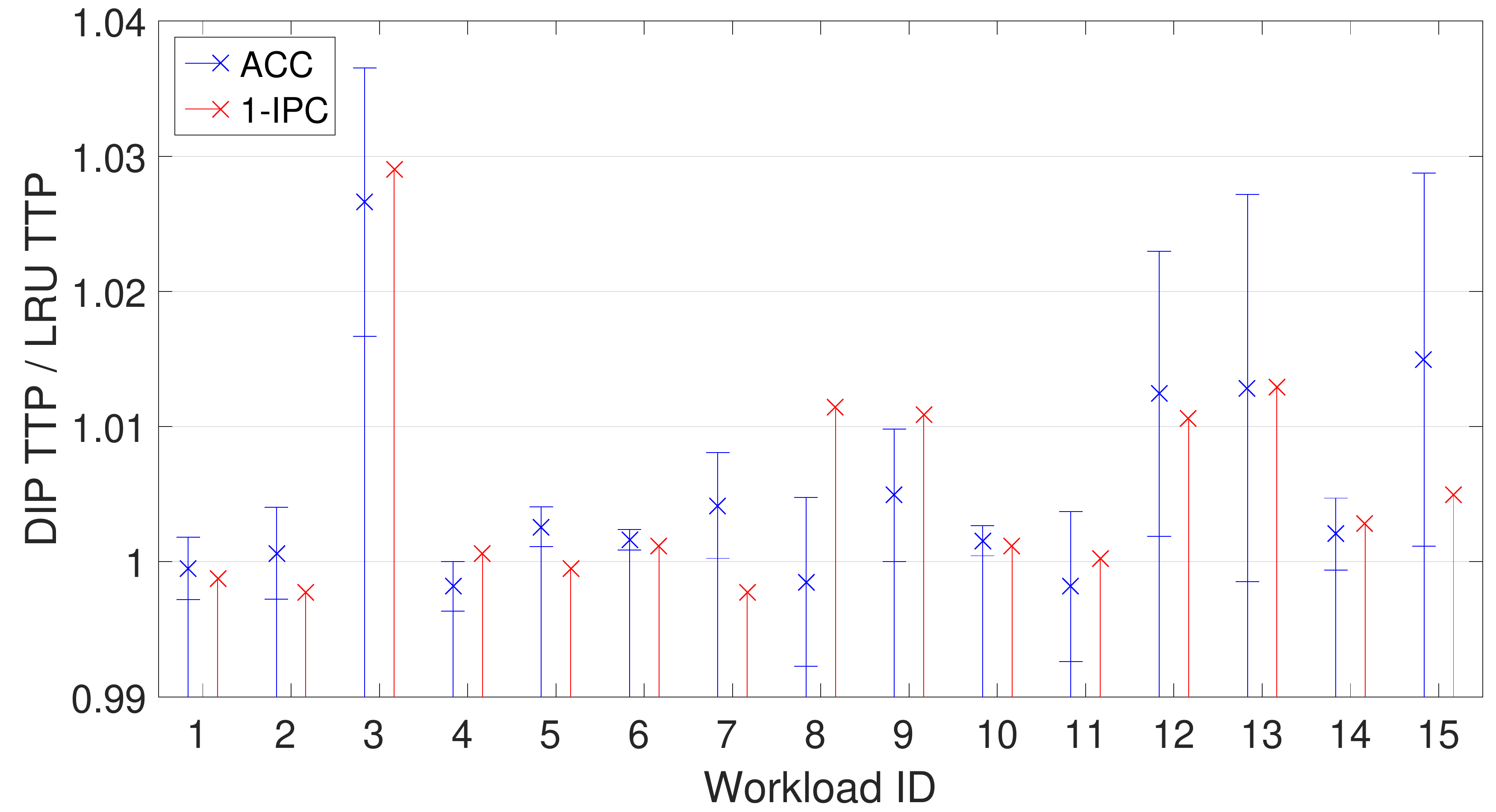}} \\
	\subfloat[Using \WSU{} for comparison \label{fig: wsu dip/lru}]{\includegraphics[width=0.45\textwidth]{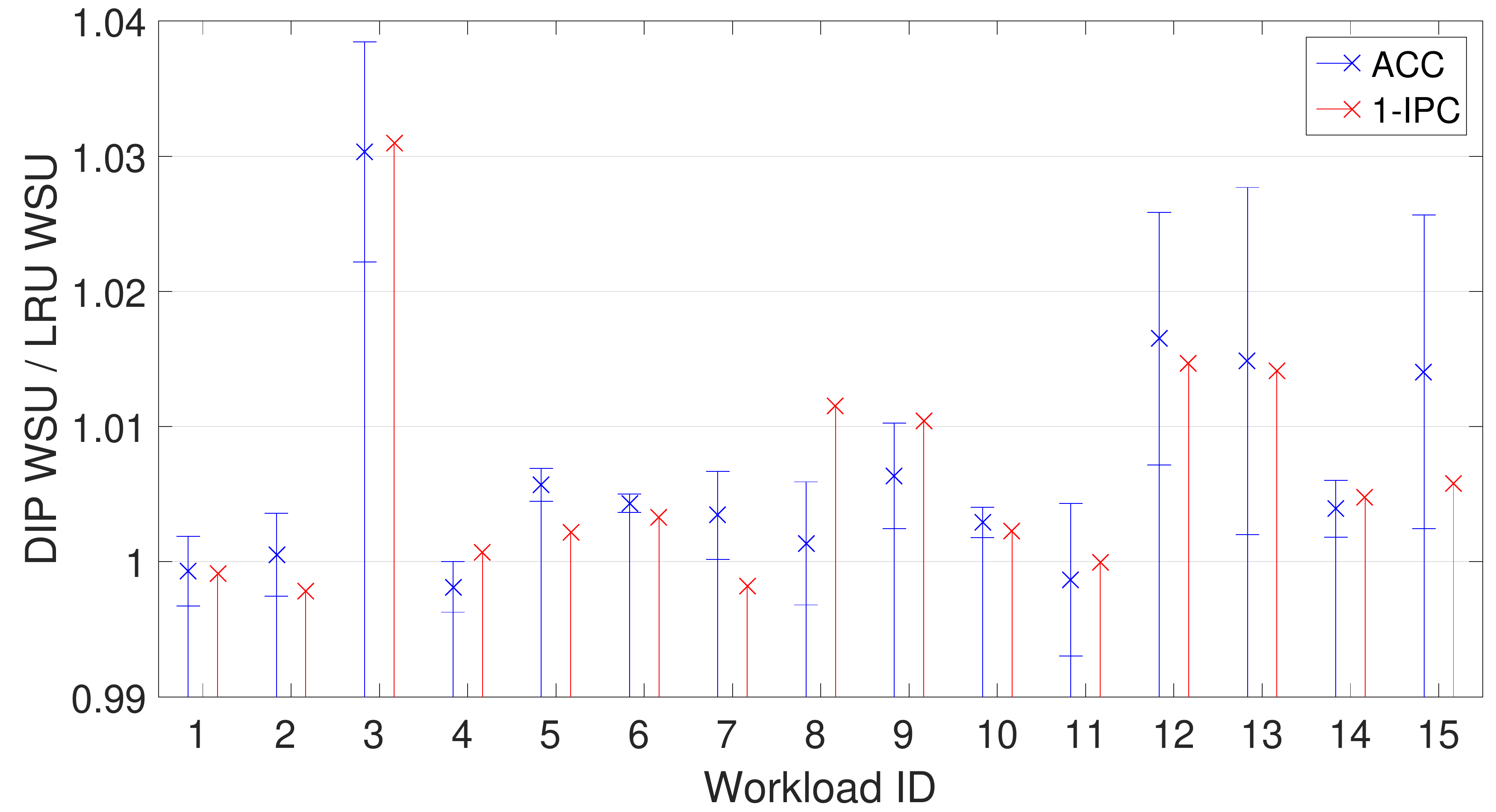}}
	\caption{Comparison of DIP versus LRU (baseline) on ACC and \OneIPC{} models} \label{fig: dip/lru}
\end{figure}

\begin{figure}[!htb]
	\centering
	\subfloat[Using MPKI for comparison \label{fig: mpki drrip/dip}]{\includegraphics[width=0.45\textwidth]{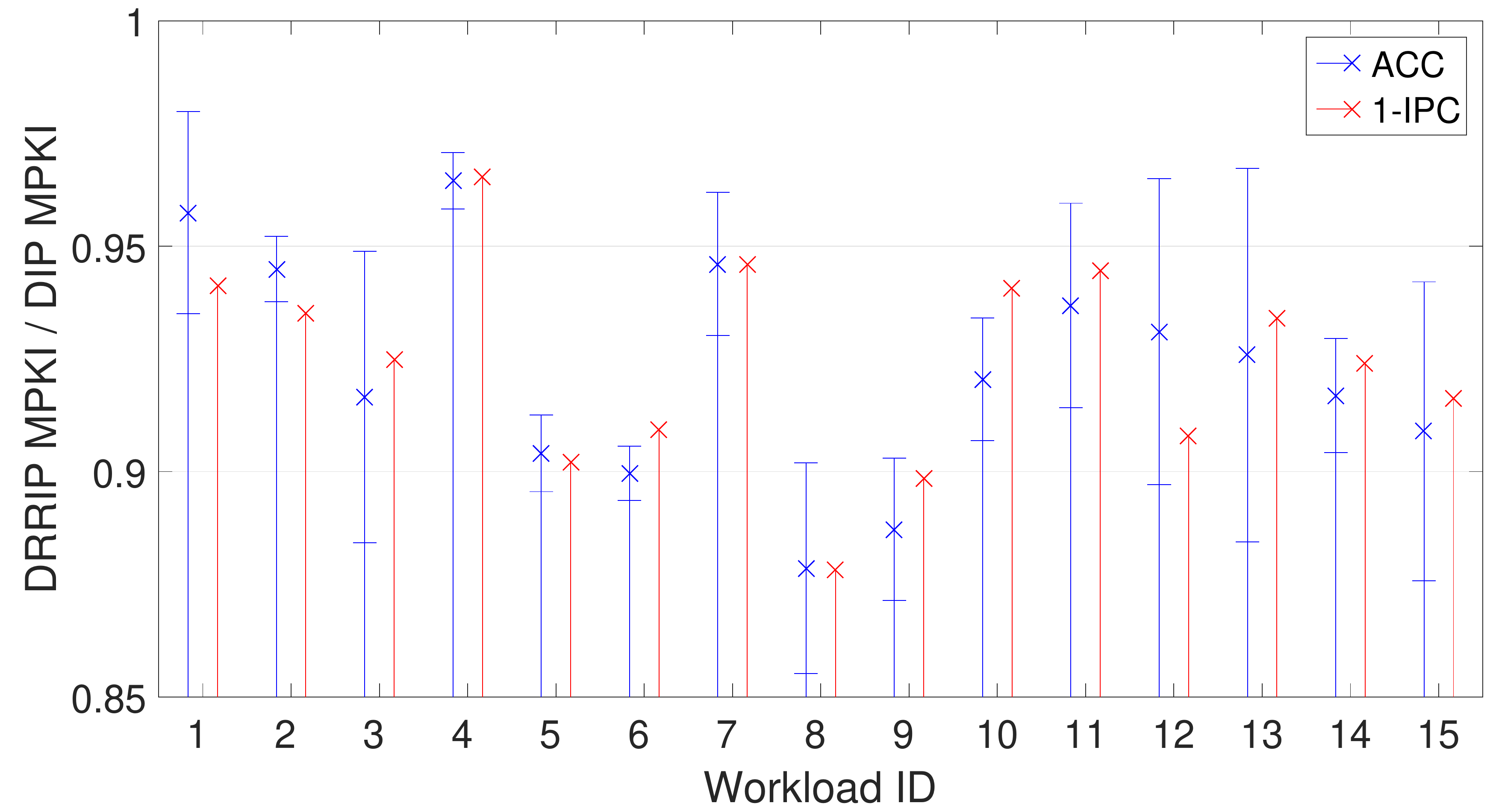}} \\
	\subfloat[Using \TTP{} for comparison \label{fig: ipc drrip/dip}]{\includegraphics[width=0.45\textwidth]{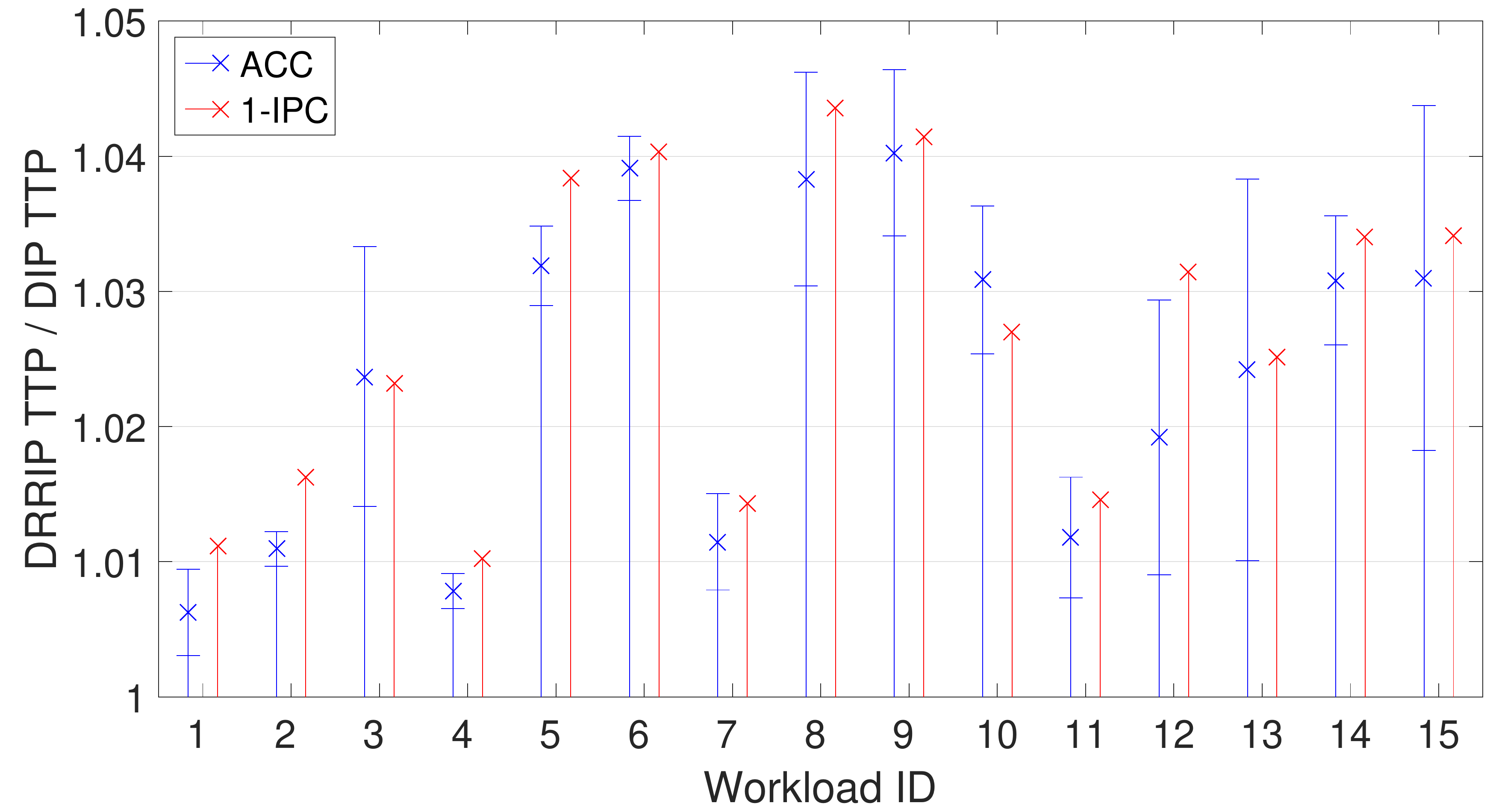}} \\
	\subfloat[Using \WSU{} for comparison \label{fig: wsu drrip/dip}]{\includegraphics[width=0.45\textwidth]{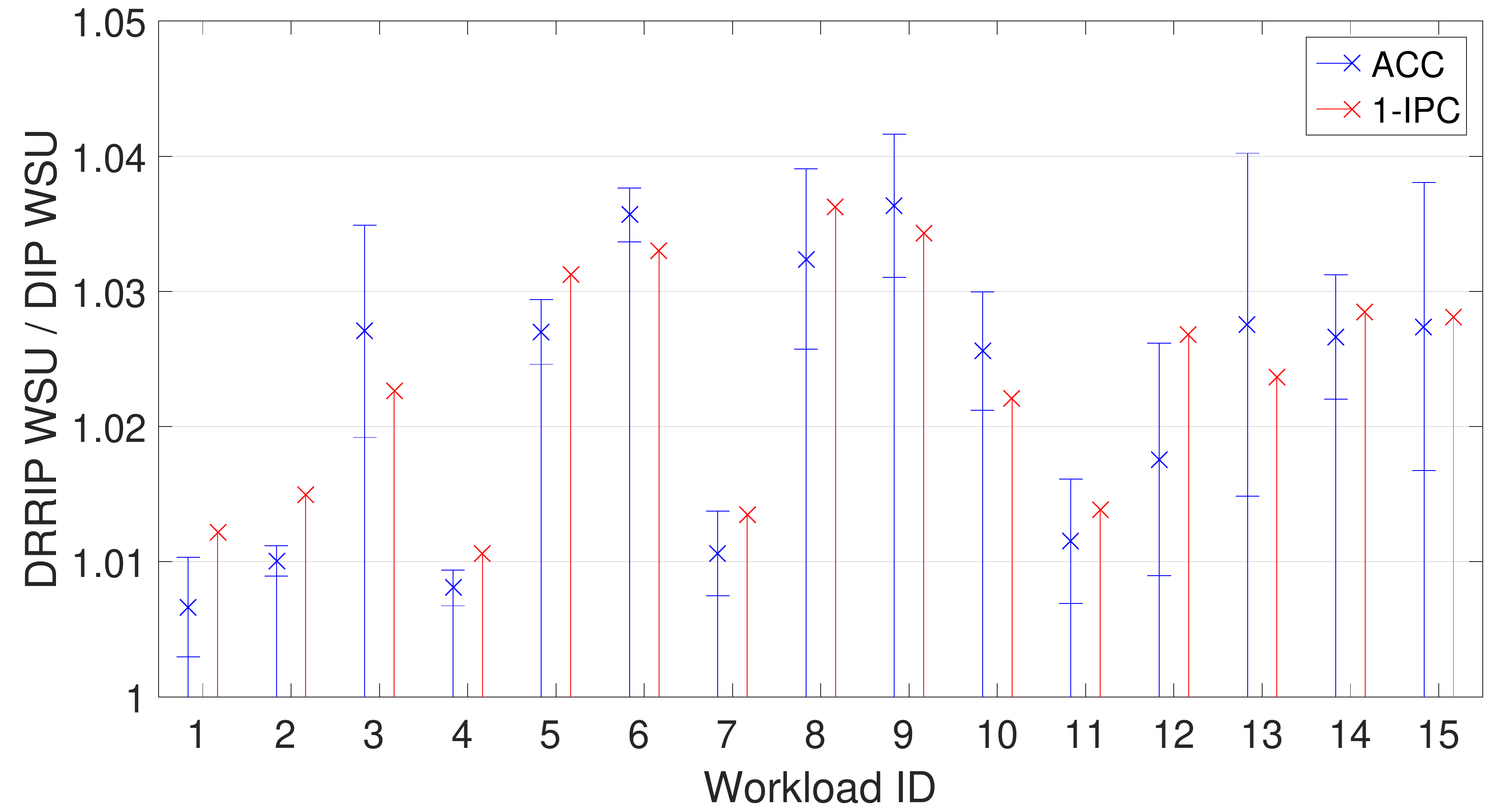}}
	\caption{Comparison of DRRIP versus DIP (baseline) on ACC and \OneIPC{} models} \label{fig: drrip/dip}
\end{figure}

Figures \ref{fig: dip/lru}$\sim$\ref{fig: drrip/dip} show the improvement ratios (the cross markers) calculated using mean metric values in Figure \ref{fig: abs mean result} for each workload, when comparing each pair of replacement policies using different metrics on both ACC and \OneIPC{} models.
The variation range (\ie{} $\mu\pm 1.28\sigma$) of each ratio on the ACC model is illustrated by the interval between horizontal bars in Figures \ref{fig: dip/lru}$\sim$\ref{fig: drrip/dip}.

\begin{figure}[!htb]
	\centering
	\subfloat[Using MPKI for comparison \label{fig: mpki drrip/lru}]{\includegraphics[width=0.45\textwidth]{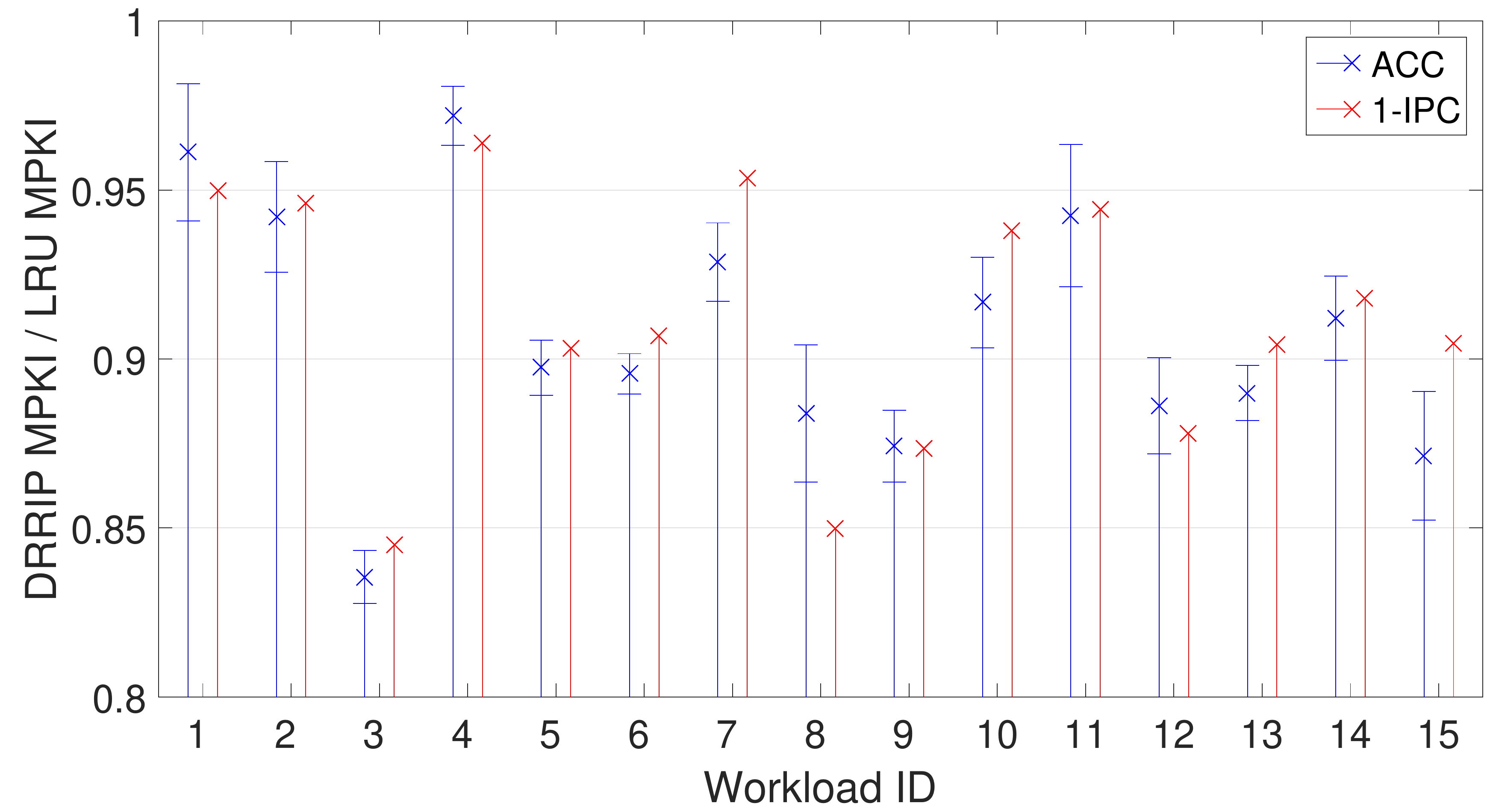}} \\
	\subfloat[Using \TTP{} for comparison \label{fig: ipc drrip/lru}]{\includegraphics[width=0.45\textwidth]{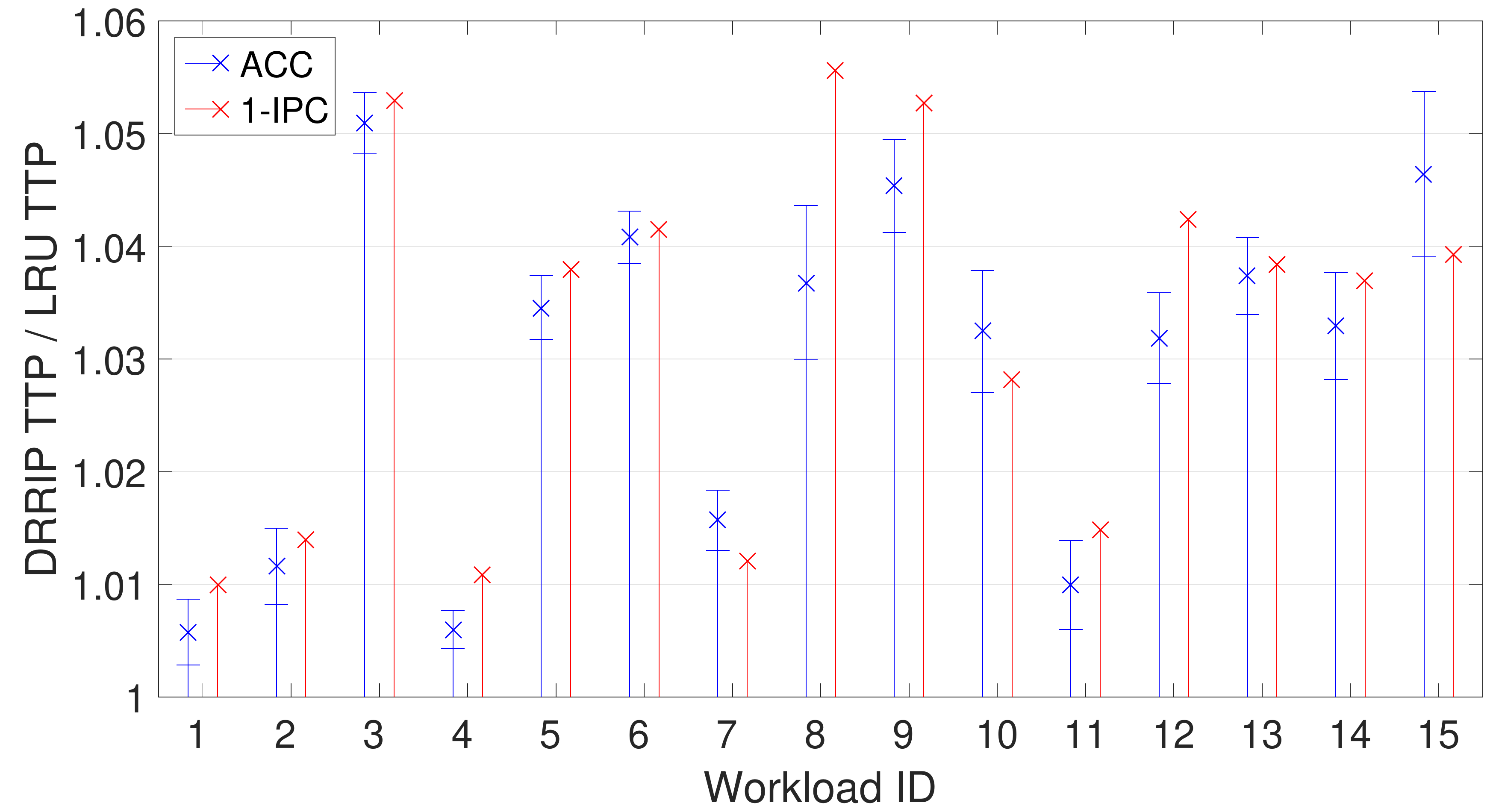}} \\
	\subfloat[Using \WSU{} for comparison \label{fig: wsu drrip/lru}]{\includegraphics[width=0.45\textwidth]{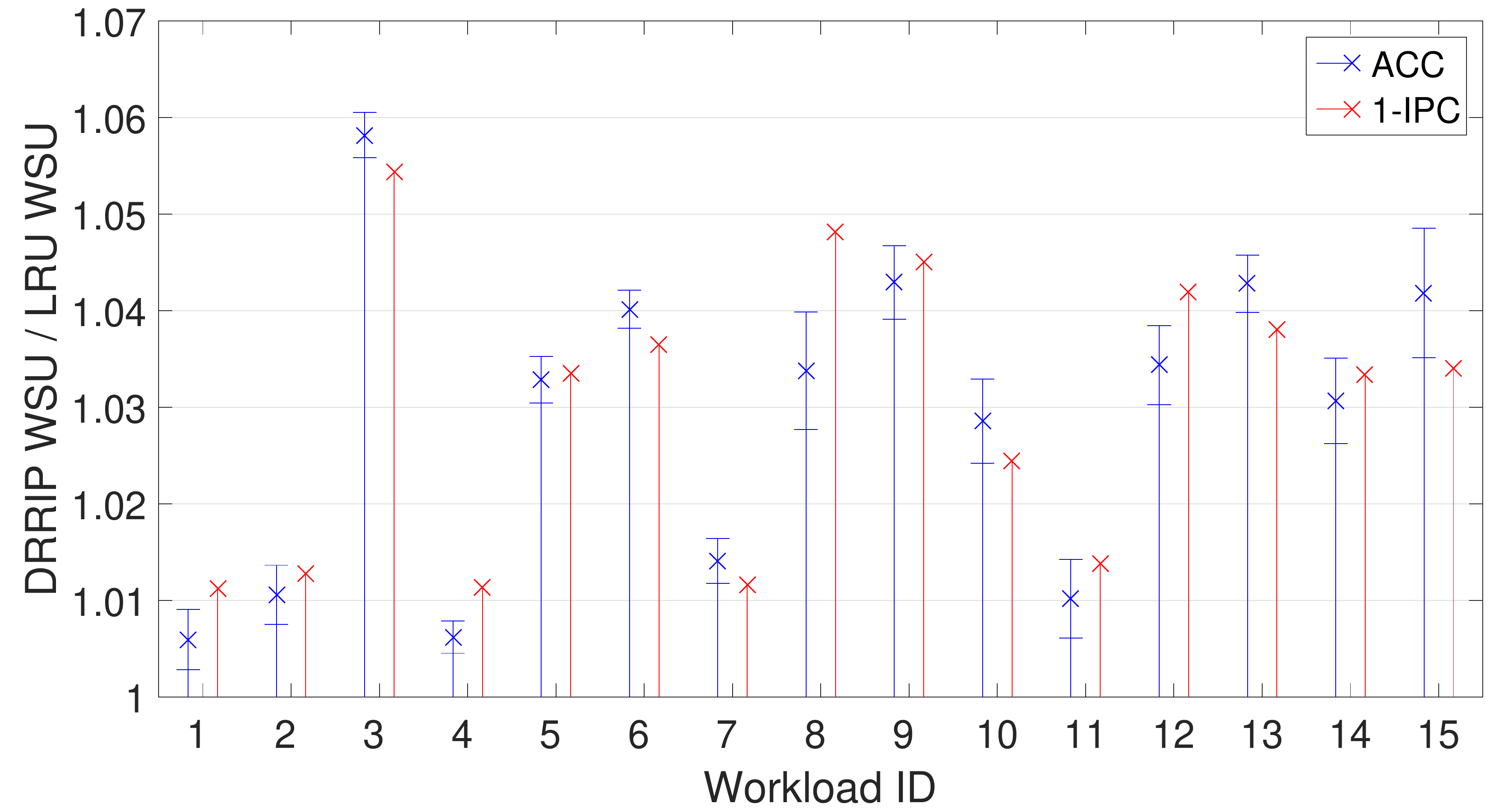}}
	\caption{Comparison of DRRIP versus LRU (baseline) on ACC and \OneIPC{} models} \label{fig: drrip/lru}
\end{figure}

Table~\ref{tab: mismatch worklaod id} shows the IDs of workloads that the decisions from \OneIPC{} model fail to match the decisions from ACC model for each pair of policies -- that is, the improvement ratios of ACC and \OneIPC{} models are on opposite sides of 1.
Note that mismatches on the decisions only happen in the comparison of DIP against LRU, so we do not list the other two comparisons in the table.

\begin{table}[!htb]
\centering
\footnotesize
\begin{tabular}{|c|c|c|c|}
	\hline
	& MPKI as metric & \TTP{} as metric & \WSU{} as metric \\
	\hline
	DIP vs. LRU & 2,4,5,7,8,11 & 2,4,5,7,8,11  & 2,4,7 \\
	\hline
\end{tabular}
\caption{Workload IDs that the comparison results from the \OneIPC{} model do not match those of the ACC model} \label{tab: mismatch worklaod id}
\end{table}

We notice that the \OneIPC{} model matches the ACC model exactly in the comparisons of DRRIP versus LRU and DRRIP versus DIP, but there are several mismatches when comparing DIP against LRU. 
If we look at Figure~\ref{fig: dip/lru} in more detail, we find that the improvement of DIP over LRU is not obvious.
Furthermore, the variation ranges imply that the wrong decisions derived from the \OneIPC{} model may also be drawn if one only conducts the experiment once even using the ACC model.

For example in Figure \ref{fig: mpki dip/lru}, among the six workloads that exhibit mismatches when using MPKI as metric, the variation ranges of the ACC model for four of them (ID: 2, 4, 8, 11) cross the ratio line corresponding to 1.
Namely for each of these four workloads, if one conducts a single experiment using the ACC model, he/she will have at least 10\% probability to make the same qualitative  decisions as those implied by the \OneIPC{} model.
Similar situations happen for workloads 2, 8 and 11 in Figure~\ref{fig: ipc dip/lru} when using \TTP{} as metric, and for workloads 2 and 4 in Figure~\ref{fig: wsu dip/lru} when using \WSU{} as metric. 

Workloads that exhibit \emph{clear} mismatches (\ie{} excluding workloads that variation ranges of the ACC model cross the ratio line corresponding to 1) only exist in the comparison of DIP against LRU, and are shown in Table \ref{tab: clear mismatch workload id}.
Considering that we only have 6 clear mismatches among 135 comparisons (3 metrics$\times$3 pairs for comparison$\times$15 workloads), \OneIPC{} model is qualitatively quite accurate. 
Furthermore, since \OneIPC{} model exactly agrees with ACC model on the clear improvement of DRRIP over the other policies while the advantage of DIP over LRU is insignificant, we may conclude that \OneIPC{} model is qualitatively accurate in showing a clear improvement while it may fail to match the ACC model when comparing polices that yield similar performance.

\begin{table}
	\centering
	\footnotesize
	\begin{tabular}{|c|c|c|c|}
		\hline
		& MPKI as metric & \TTP{} as metric & \WSU{} as metric \\ \hline
		DIP vs. LRU & 5,7 & 4,5,7 & 7 \\ \hline
	\end{tabular}
	\caption{Workloads IDs that exhibit {clear} mismatches} \label{tab: clear mismatch workload id}
\end{table}

\noindent\textbf{Quantifying the impact of using \OneIPC{} model:}
In order to quantify the impact of using the \OneIPC{} model instead of the ACC model, we compare the differences between two models against the variation of ACC model in terms of the geometric means of improvement rations.
We choose to study the geometric mean, because we can simply compare it to 1 and determine the better policy.

We can also estimate the variation of the geometric mean on the ACC model using a similar way that we have used for the improvement ratio for each single workload.
Assume that $r_i$ ($i=1\ldots 15$) is the random variable of the improvement ratio of certain metric for workload $i$ when comparing two policies on the ACC model, and that $g$ is the geometric mean over all $r_i$.
We have shown earlier that $r_i$ approximately follows a normal distribution $N(\mu(r_i),\sigma(r_i))$, and we have calculated $\mu(r_i)$ and $\sigma(r_i)$.
Since $\sigma(r_i)$ is also much smaller than $\mu(r_i)$, we can again do Taylor expansion on the geometric mean ($\delta(r_i)$ represents $r_i - \mu(r_i)$):
\begin{eqnarray}
g & = & \sqrt[15]{\prod_{i=1}^{15} r_i} = \sqrt[15]{\prod_{i=1}^{15}\mu(r_i)} \cdot \sqrt[15]{\prod_{i=1}^{15} \left(1 + \frac{\delta(r_i)}{\mu(r_i)}\right)} \nonumber \\
& = & \sqrt[15]{\prod_{i=1}^{15}\mu(r_i)} \cdot \left( 1 + \frac{1}{15} \sum_{i=1}^{15} \frac{\delta(r_i)}{r_i} \right) \nonumber
\end{eqnarray}
Therefore the geometric mean $g$ of improvement ratios also approximately follows a normal distribution $N(\mu(g),\sigma(g))$, where $\mu(g) = \sqrt[15]{\prod_{i=1}^{15} \mu(r_i)}$ and 
\begin{displaymath}
\sigma(g) = \frac{\mu(g)}{15} \cdot \sqrt{\sum_{i=1}^{15}\left( \frac{\sigma(r_i)}{\mu(r_i)} \right)^2}.
\end{displaymath}

\begin{figure}[!htb]
	\centering
	\subfloat[Using MPKI for comparison]{\includegraphics[width=0.48\columnwidth]{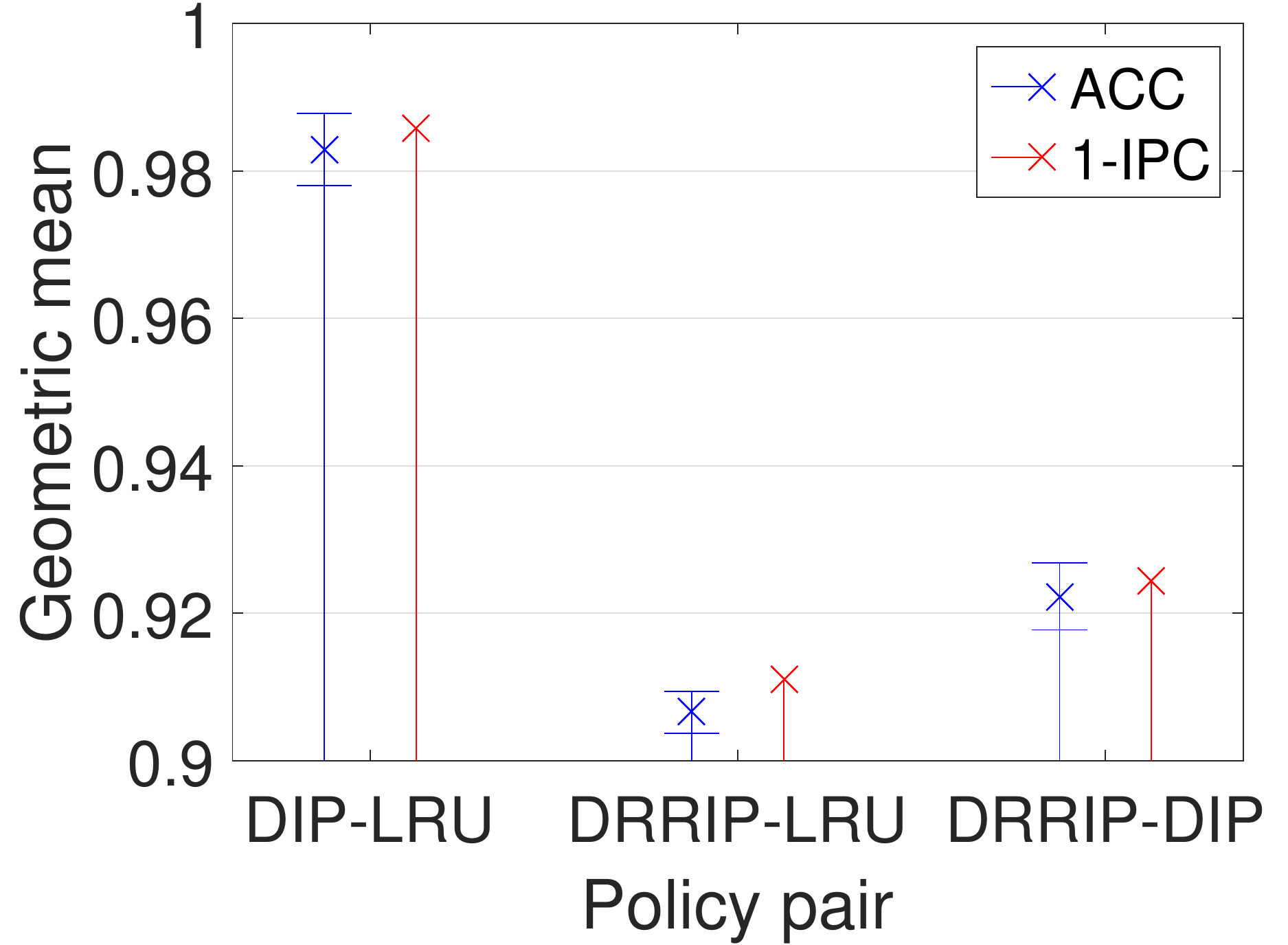}} \hspace{5pt}
	\subfloat[Using \TTP{} for comparison]{\includegraphics[width=0.48\columnwidth]{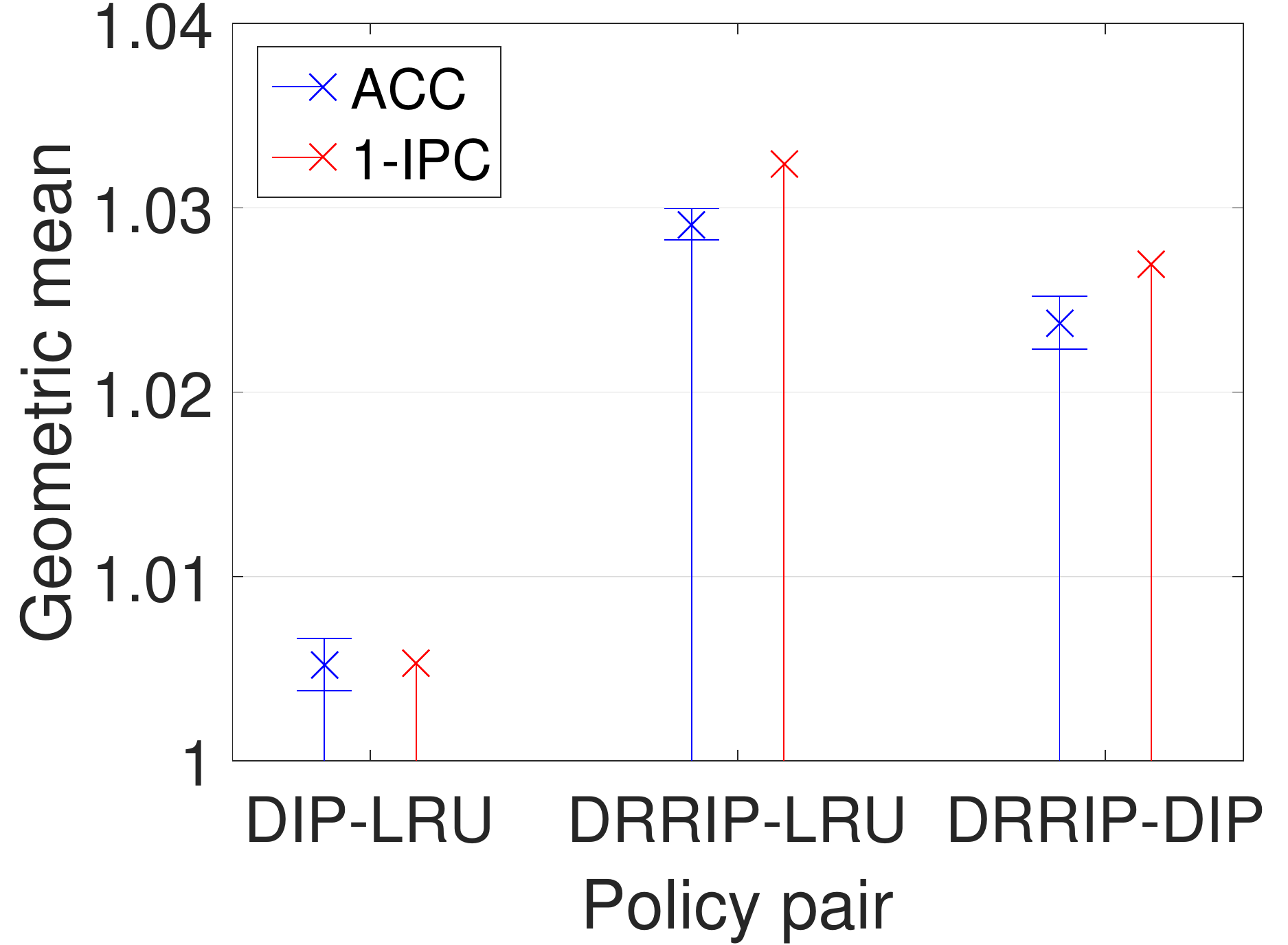}} \\
	\subfloat[Using \WSU{} for comparison]{\includegraphics[width=0.48\columnwidth]{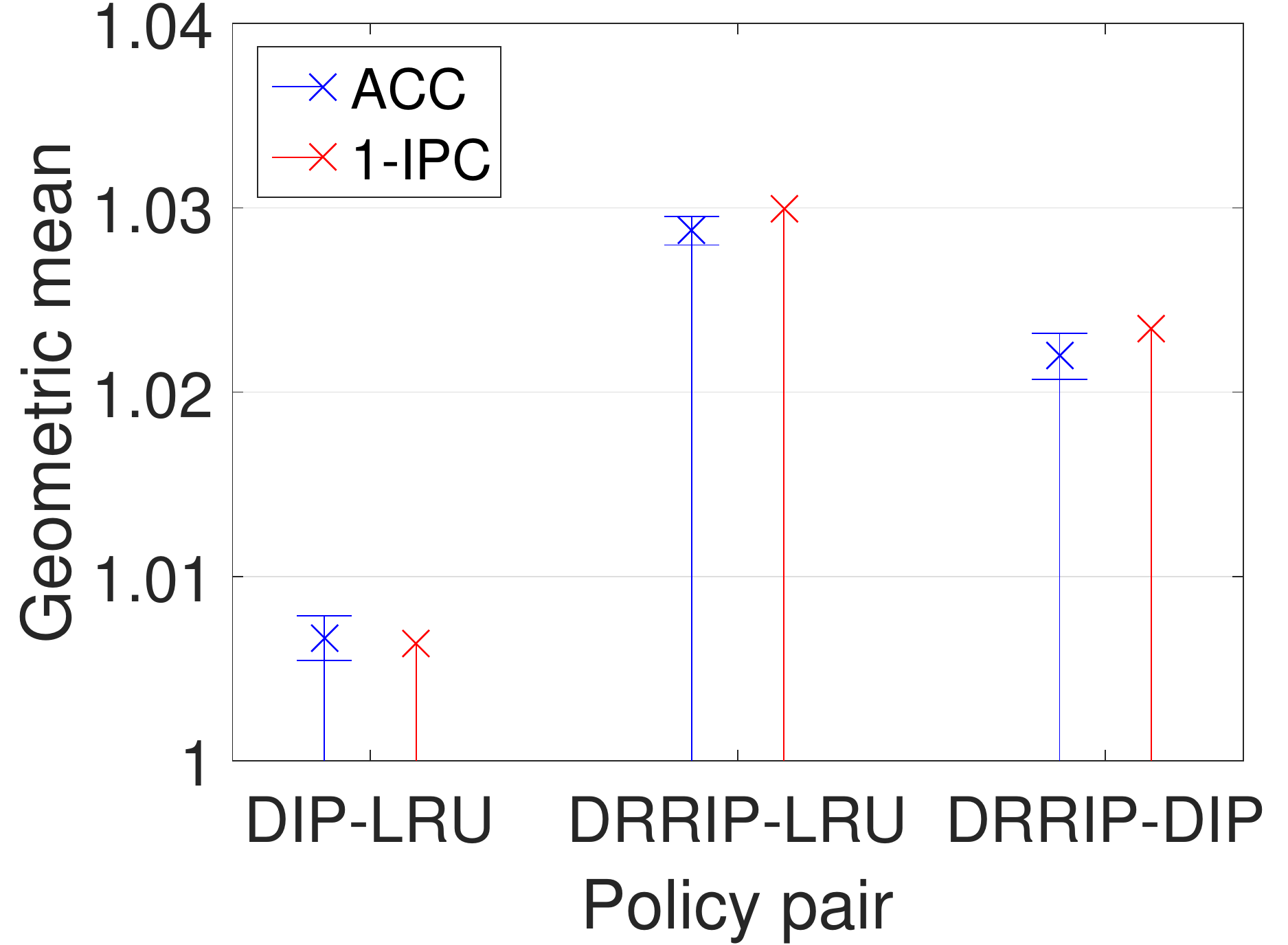}}
	\caption{Geometric means of improvement ratios for each policy pair using three metrics on ACC and \OneIPC{} models} \label{fig: geo mean}
\end{figure}

Figure \ref{fig: geo mean} shows the geometric means of improvement ratios (the cross markers) calculated using mean metric values in Figure \ref{fig: abs mean result} for each metric when comparing each pair of replacement policy on both ACC and \OneIPC{} models.
The standard deviation (\ie{} $\sigma$) of the normal distribution of each geometric mean on the ACC model is illustrated by the distance from the horizontal bar to the cross marker in Figure \ref{fig: geo mean}.
As we can see from the figure, the differences between the \OneIPC{} geometric means and the corresponding ACC values are at the same order of magnitude of the standard deviations of the ACC model.
And sometimes the \OneIPC{} results even fall into the variation ranges of the ACC model.
These observations all imply that the error induced by using \OneIPC{} model is comparable to the variation of ACC model.

\noindent\textbf{Summary:} 
We first showed that using the \OneIPC{} model instead of the ACC model did not significantly change the characteristics of the workloads.
We then illustrated that the \OneIPC{} model could give qualitatively accurate results in the evaluation of different polices when the improvement is unambiguous. 
Furthermore, we demonstrated that the quantitative impact of using the \OneIPC{} model to compare LLC policies was at the same order as the impact of run-to-run variation when the cycle-accurate experiment cannot be run multiple times.

\section{Study 2: Scalability of Multithread Benchmarks} \label{sec: scale}

In this section, we evaluate the scalability of several multithread benchmarks using ACC and \OneIPC{} models with up to 16 cores.

\subsection{Simulator Improvement and Settings}
We implemented 1-, 2-, 4-, 8-, and 16-core systems with 512KB, 1MB, 2MB, 4MB and 8MB L2 caches respectively.
All other parameters for the core model and main memory are the same as Table~\ref{tab: base-setup}.

These simulators were each implemented on a single VC707 board, and for systems larger than the 4-core system, it becomes a challenge to implement these larger systems in the framework of Arete.
This is because Arete translates each hardware module in the processor into a \LIBDN{} node, which takes multiple cycles to simulate the behavior of the original module in one cycle.
Therefore when Arete models a multicore system with $N$ cores, it replicates the \LIBDN{} nodes of the core model for $N$ times.
Due to the resource constraints of the VC707 board, it is impossible to fit more than four core models on a single FPGA.
The solution Arete employs is to map the design to a multi-FPGA board. 
Unfortunately, we do not have a multi-FPGA board. 
Handling the hardware for inter-FPGA communication requires considerate engineering work, and the inter-board communication latency ends up being high.

Our solution is to apply fine-grained time-division multiplexing \cite{pellauer2011hasim} to Arete's \LIBDN{}s.
Namely one \LIBDN{} node will model the same circuit module in multiple cores by simulating the functionality of each core's module one by one.
The design avoids deadlock due to the absence of combinational paths between cores in the target architecture.
In this way, we can save FPGA resources because the logic for modeling functionality inside the core can be reused.
One thing we do have to add for time-division multiplexing is registers for each state in the target architecture -- such as the PC, registers in the pipeline FIFOs, \emph{etc.} -- so a single state register will be expanded to a vector of registers, each of which corresponds to the state register in one core.

\subsection{Multithread Benchmarks}
We choose 6 multithread benchmarks from the PARSEC-3.0 \cite{bienia2008parsec} and SPLASH-2x \cite{woo1995splash}\cite{splash2x} benchmark suites, as shown in Table \ref{tab: multithread benchmarks}.
Each benchmark is run to completion with the simsmall input size.

\begin{table}[!htb]
\centering
\footnotesize
\begin{tabular}{|c|c|}
\hline
Benchmark suite & Benchmark name \\
\hline
PARSEC-3.0 & blackscholes, canneal, fluidanimate, streamcluster \\
\hline
SPLASH-2x & fft, water\_nsquared \\
\hline
\end{tabular}
\caption{Multithread benchmarks for evaluating scalability} \label{tab: multithread benchmarks}
\end{table}

\subsection{Results and Analysis}
\begin{figure}[!htb]
\centering
\includegraphics[width=0.45\textwidth]{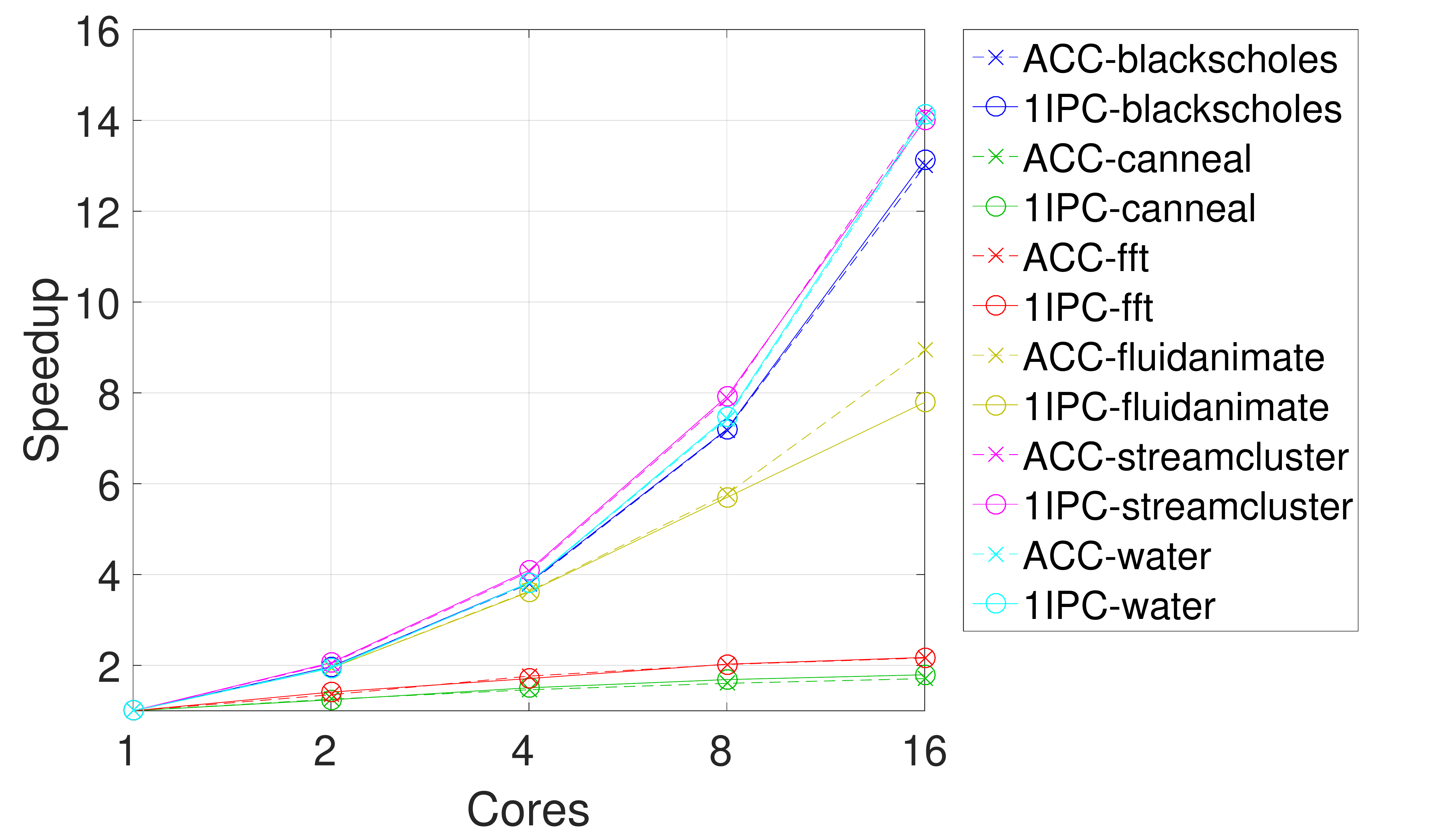}
\caption{Scalability of 5 multithread benchmarks on up to 16 cores} \label{fig: scale}
\end{figure}

We ran each of the 6 benchmarks to completion with 1, 2, 4, 8 and 16 cores, and we measured the execution time to calculate the speed up normalized to single-core performance for each model.
Figure~\ref{fig: scale} shows the scalability of each benchmark measured from the ACC core model and the 1IPC core model.

We observe that almost all of the scalability curves for the \OneIPC{} model match the corresponding curves from the ACC model perfectly except for the fluidanimate benchmark running on 16 cores. 
Currently we haven't found the reason for this single mismatch and we are still investigating it.

One interesting observation is that in our experiment the \OneIPC{} model is able to accurately capture the scalability of the water\_nsquare benchmark while Carlson \emph{et al.} report in \cite{carlson2011sniper} that the \OneIPC{} model fails to capture the scalability of this benchmark on multicore systems with out-of-order cores.
We believe this is because our ACC core model is an in-order pipeline, significantly different from an out-of order core.

\section{Study 3: Branch Predictor} \label{sec: brpred}

In this section, we evaluate the following three branch predictors on a single core using the \OneIPC{} and ACC models:
\begin{enumerate}
	\item the \textbf{tournament} branch predictor \cite{kessler1998alpha} from Alpha 21264;
	\item the \textbf{path-based neural} branch predictor \cite{jimenez2003fast} with ahead pipelining; and
	\item the \textbf{TAGE} (TAgged GEometric history length) branch predictor \cite{Andre2006Tage} based on Seznec's source code \cite{Andre2014tagescl} submitted to CBP-4 (Championship Branch Prediction).
\end{enumerate}
The sizes of the storage used by three branch predictors are all around 4KB.

\subsection{Benchmarks and Measurement}
We evaluate these three branch predictors on a single core with 512KB L2 cache using 5 benchmarks from SPEC CINT2006 benchmark suite, which could incur high misprediction rates, as shown in Table~\ref{tab: brpred benchmarks}. 
All benchmarks are run to completion with the test input and we measure the misprediction rate only for conditional branches (excluding indirect jumps).

\begin{table}[!htb]
\centering
\begin{footnotesize}
\begin{tabular}{|c|c|}
\hline
Benchmark suite & Benchmark name \\
\hline
SPEC CINT2006 & gobmk, hmmer, mcf, omnetpp, sjeng \\
\hline
\end{tabular}
\end{footnotesize}
\caption{Benchmarks for evaluating branch predictors} \label{tab: brpred benchmarks}
\end{table}

\subsection{Results and Analysis}
\begin{figure}[!htb]
\centering
\includegraphics[width=0.43\textwidth]{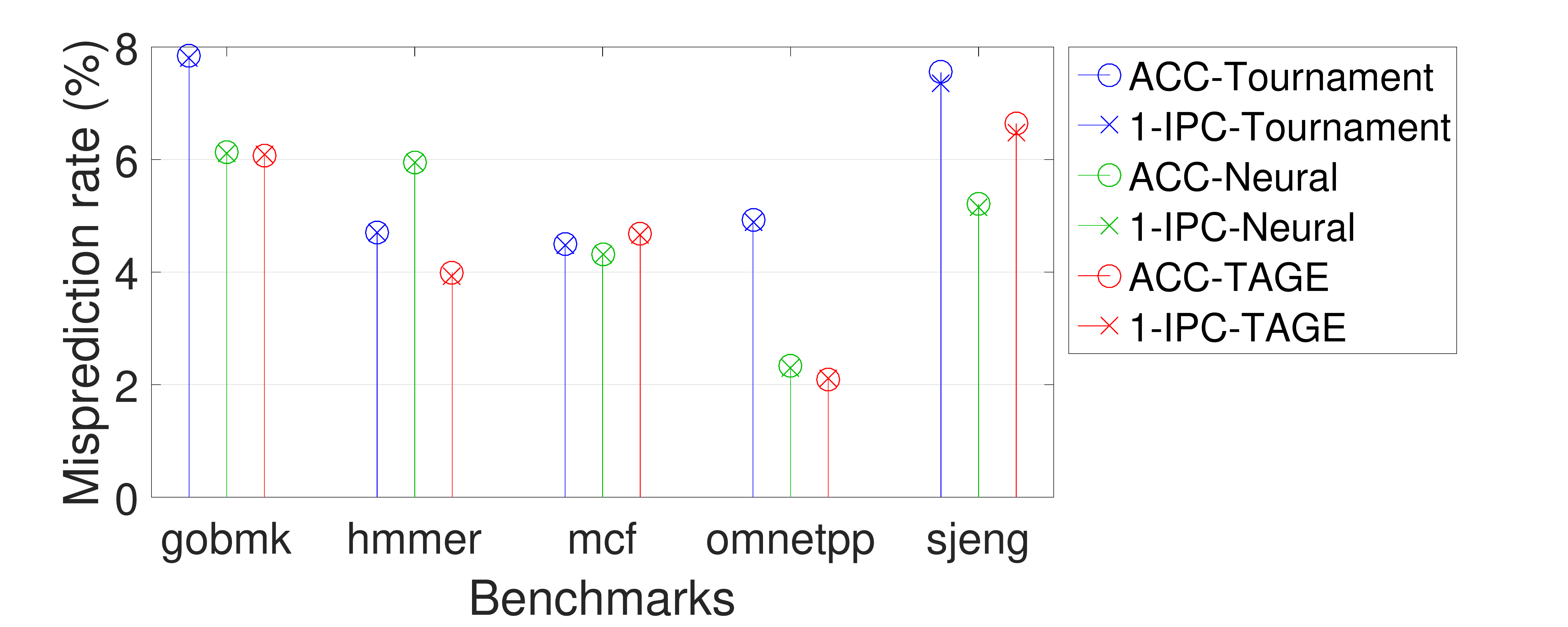}
\caption{Misprediction rates for all three branch predictors on ACC and \OneIPC{} models} \label{fig: mispred rate}
\end{figure}

Figure~\ref{fig: mispred rate} shows the misprediction rates for all three branch predictors measured on the ACC and \OneIPC{} models.
The results from the \OneIPC{} model matches those from the ACC model extremely well. 
We believe the similarity is because there are only three pipeline stages between branch prediction and resolution, and the probability of multiple outstanding unresolved branches is quite low. 
Therefore, the \OneIPC{} model -- which can resolve branch in the same cycle as making prediction -- is not significantly different from the ACC model in terms of training the branch predictor. 

\section{Conclusion} \label{sec: conclude}

Simplifications are used often in computer architectural simulation to reduce the simulation time.
By running a cycle-accurate full-system simulator side-by-side with a version of the same simulator that uses a \OneIPC{} core model, we are able to isolate and measure the effects of the \OneIPC{} core model simplification.
We find that, although the \OneIPC{} model does not report accurate absolute metric values, the relative behavior of the \OneIPC{} model matches that of the ACC model.

First, by normalizing metric results across 15 workloads for three LLC cache replacement policies, we showed that the \OneIPC{} core model does not distort the characteristics of the results across the workloads for each replacement policy.
Further exploring the LLC experiment results we showed that using the \OneIPC{} core model to make comparisons between replacement policies resulted in the correct comparisons most of the times, and when the comparisons were not correct, it was often due to brittle policies.
Third, by running multicore benchmarks on various numbers of cores, we showed that the \OneIPC{} model appropriately matches the scaling trends shown displayed by the ACC model.
Finally, by comparing three branch predictors across the two models, we showed that the \OneIPC{} model matched the branch prediction accuracy of the ACC model.

We find that the simplified \OneIPC{} core model is useful to produce qualitative comparisons between architectural configurations, but this is not a suggestion to ignore cycle-accurate models in favor of simplified \OneIPC{} models.
This is merely an invitation to design simplified core models in parallel with cycle accurate models to prove their usefulness before switching to the simplified core model for experiments.

\section{Acknowledgment}
We thank Asif Khan for his pioneering work in studying the problem of simulation accuracy, and his great help in using the Arete simulator.


\bibliographystyle{ieeetr}
\bibliography{ref}

\begin{thebibliography}{10}

\bibitem{khan2012fast}
A.~Khan, M.~Vijayaraghavan, S.~Boyd-Wickizer, {\em et~al.}, ``Fast and
  cycle-accurate modeling of a multicore processor,'' in {\em Performance
  Analysis of Systems and Software (ISPASS), 2012 IEEE International Symposium
  on}, pp.~178--187, IEEE, 2012.

\bibitem{martin2005multifacet}
M.~M. Martin, D.~J. Sorin, B.~M. Beckmann, M.~R. Marty, M.~Xu, A.~R.
  Alameldeen, K.~E. Moore, M.~D. Hill, and D.~A. Wood, ``Multifacet's general
  execution-driven multiprocessor simulator (gems) toolset,'' {\em ACM SIGARCH
  Computer Architecture News}, vol.~33, no.~4, pp.~92--99, 2005.

\bibitem{binkert2006m5}
N.~L. Binkert, R.~G. Dreslinski, L.~R. Hsu, K.~T. Lim, A.~G. Saidi, and S.~K.
  Reinhardt, ``The m5 simulator: Modeling networked systems,'' {\em IEEE
  Micro}, vol.~26, no.~4, pp.~52--60, 2006.

\bibitem{patel2011marss}
A.~Patel, F.~Afram, S.~Chen, and K.~Ghose, ``Marss: a full system simulator for
  multicore x86 cpus,'' in {\em Proceedings of the 48th Design Automation
  Conference}, pp.~1050--1055, ACM, 2011.

\bibitem{argollo2009cotson}
E.~Argollo, A.~Falc{\'o}n, P.~Faraboschi, M.~Monchiero, and D.~Ortega,
  ``Cotson: infrastructure for full system simulation,'' {\em ACM SIGOPS
  Operating Systems Review}, vol.~43, no.~1, pp.~52--61, 2009.

\bibitem{jaleel2008cmp}
A.~Jaleel, R.~S. Cohn, C.-K. Luk, and B.~Jacob, ``Cmp\$im: A pin-based
  on-the-fly multi-core cache simulator,'' in {\em Proceedings of the Fourth
  Annual Workshop on Modeling, Benchmarking and Simulation (MoBS), co-located
  with ISCA}, pp.~28--36, 2008.

\bibitem{miller2010graphite}
J.~E. Miller, H.~Kasture, G.~Kurian, C.~Gruenwald, N.~Beckmann, C.~Celio,
  J.~Eastep, and A.~Agarwal, ``Graphite: A distributed parallel simulator for
  multicores,'' in {\em High Performance Computer Architecture (HPCA), 2010
  IEEE 16th International Symposium on}, pp.~1--12, IEEE, 2010.

\bibitem{carlson2011sniper}
T.~E. Carlson, W.~Heirman, and L.~Eeckhout, ``Sniper: exploring the level of
  abstraction for scalable and accurate parallel multi-core simulation,'' in
  {\em Proceedings of 2011 International Conference for High Performance
  Computing, Networking, Storage and Analysis}, p.~52, ACM, 2011.

\bibitem{sanchez2013zsim}
D.~Sanchez and C.~Kozyrakis, ``Zsim: fast and accurate microarchitectural
  simulation of thousand-core systems,'' in {\em Proceedings of the 40th Annual
  International Symposium on Computer Architecture}, pp.~475--486, ACM, 2013.

\bibitem{luk2005pin}
C.-K. Luk, R.~Cohn, R.~Muth, H.~Patil, A.~Klauser, G.~Lowney, S.~Wallace, V.~J.
  Reddi, and K.~Hazelwood, ``Pin: building customized program analysis tools
  with dynamic instrumentation,'' {\em ACM Sigplan Notices}, vol.~40, no.~6,
  pp.~190--200, 2005.

\bibitem{genbrugge2010interval}
D.~Genbrugge, S.~Eyerman, and L.~Eeckhout, ``Interval simulation: Raising the
  level of abstraction in architectural simulation,'' in {\em High Performance
  Computer Architecture (HPCA), 2010 IEEE 16th International Symposium on},
  pp.~1--12, IEEE, 2010.

\bibitem{nowatzki2014harmful}
T.~Nowatzki, J.~Menon, C.-H. Ho, and K.~Sankaralingam, ``gem5, gpgpusim, mcpat,
  gpuwattch, "your favorite simulator here" considered harmful,'' in {\em 11th
  Annual Workshop on Duplicating, Deconstructing and Debunking}, 2014.

\bibitem{gibson2000flash}
J.~Gibson, R.~Kunz, D.~Ofelt, M.~Horowitz, J.~Hennessy, and M.~Heinrich,
  ``Flash vs.(simulated) flash: Closing the simulation loop,'' in {\em ACM
  SIGARCH Computer Architecture News}, vol.~28, pp.~49--58, ACM, 2000.

\bibitem{desikan2001measuring}
R.~Desikan, D.~Burger, and S.~W. Keckler, ``Measuring experimental error in
  microprocessor simulation,'' in {\em Proceedings of the 28th annual
  international symposium on Computer architecture}, pp.~266--277, ACM, 2001.

\bibitem{cain2002precise}
H.~W. Cain, K.~M. Lepak, B.~A. Schwartz, and M.~H. Lipasti, ``Precise and
  accurate processor simulation,'' in {\em Workshop on Computer Architecture
  Evaluation using Commercial Workloads, HPCA}, vol.~8, 2002.

\bibitem{mutlu2004understanding}
O.~Mutlu, H.~Kim, D.~N. Armstrong, and Y.~N. Patt, ``Understanding the effects
  of wrong-path memory references on processor performance,'' in {\em
  Proceedings of the 3rd workshop on Memory performance issues: in conjunction
  with the 31st international symposium on computer architecture}, pp.~56--64,
  ACM, 2004.

\bibitem{sendag2006quantifying}
R.~Sendag, A.~Yilmazer, J.~J. Yi, and A.~K. Uht, ``Quantifying and reducing the
  effects of wrong-path memory references in cache-coherent multiprocessor
  systems,'' in {\em Parallel and Distributed Processing Symposium, 2006. IPDPS
  2006. 20th International}, pp.~10--pp, IEEE, 2006.

\bibitem{yi2005characterizing}
J.~J. Yi, S.~V. Kodakara, R.~Sendag, D.~J. Lilja, and D.~M. Hawkins,
  ``Characterizing and comparing prevailing simulation techniques,'' in {\em
  High-Performance Computer Architecture, 2005. HPCA-11. 11th International
  Symposium on}, pp.~266--277, IEEE, 2005.

\bibitem{sherwood2002automatically}
T.~Sherwood, E.~Perelman, G.~Hamerly, and B.~Calder, ``Automatically
  characterizing large scale program behavior,'' {\em ACM SIGARCH Computer
  Architecture News}, vol.~30, no.~5, pp.~45--57, 2002.

\bibitem{wunderlich2003smarts}
R.~E. Wunderlich, T.~F. Wenisch, B.~Falsafi, and J.~C. Hoe, ``Smarts:
  Accelerating microarchitecture simulation via rigorous statistical
  sampling,'' in {\em Computer Architecture, 2003. Proceedings. 30th Annual
  International Symposium on}, pp.~84--95, IEEE, 2003.

\bibitem{vijayaraghavan2009bounded}
M.~Vijayaraghavan {\em et~al.}, ``Bounded dataflow networks and
  latency-insensitive circuits,'' in {\em Formal Methods and Models for
  Co-Design, 2009. MEMOCODE'09. 7th IEEE/ACM International Conference on},
  pp.~171--180, IEEE, 2009.

\bibitem{kessler1998alpha}
R.~E. Kessler, E.~J. McLellan, and D.~A. Webb, ``The alpha 21264 microprocessor
  architecture,'' in {\em Computer Design: VLSI in Computers and Processors,
  1998. ICCD'98. Proceedings. International Conference on}, pp.~90--95, IEEE,
  1998.

\bibitem{jaleel2008adaptive}
A.~Jaleel, W.~Hasenplaugh, M.~Qureshi, J.~Sebot, S.~Steely~Jr, and J.~Emer,
  ``Adaptive insertion policies for managing shared caches,'' in {\em
  Proceedings of the 17th international conference on Parallel architectures
  and compilation techniques}, pp.~208--219, ACM, 2008.

\bibitem{qureshi2007adaptive}
M.~K. Qureshi, A.~Jaleel, Y.~N. Patt, S.~C. Steely, and J.~Emer, ``Adaptive
  insertion policies for high performance caching,'' in {\em ACM SIGARCH
  Computer Architecture News}, vol.~35, pp.~381--391, ACM, 2007.

\bibitem{jaleel2010high}
A.~Jaleel, K.~B. Theobald, S.~C. Steely~Jr, and J.~Emer, ``High performance
  cache replacement using re-reference interval prediction (rrip),'' in {\em
  ACM SIGARCH Computer Architecture News}, vol.~38, pp.~60--71, ACM, 2010.

\bibitem{DIS_stressmark}
``Data-intensive systems stressmark suite.''
  \url{http://www.ics.uci.edu/~amrm/hdu/DIS_Stressmark/DIS_stressmark.html}.

\bibitem{mccalpin1995survey}
J.~D. McCalpin, ``A survey of memory bandwidth and machine balance in current
  high performance computers,'' {\em IEEE TCCA Newsletter}, pp.~19--25, 1995.

\bibitem{franz2007ratios}
V.~H. Franz, ``Ratios: A short guide to confidence limits and proper use,''
  {\em arXiv preprint arXiv:0710.2024}, 2007.

\bibitem{pellauer2011hasim}
M.~Pellauer, M.~Adler, M.~Kinsy, A.~Parashar, and J.~Emer, ``Hasim: Fpga-based
  high-detail multicore simulation using time-division multiplexing,'' in {\em
  High Performance Computer Architecture (HPCA), 2011 IEEE 17th International
  Symposium on}, pp.~406--417, IEEE, 2011.

\bibitem{bienia2008parsec}
C.~Bienia, S.~Kumar, J.~P. Singh, and K.~Li, ``The parsec benchmark suite:
  Characterization and architectural implications,'' in {\em Proceedings of the
  17th international conference on Parallel architectures and compilation
  techniques}, pp.~72--81, ACM, 2008.

\bibitem{woo1995splash}
S.~C. Woo, M.~Ohara, E.~Torrie, J.~P. Singh, and A.~Gupta, ``The splash-2
  programs: Characterization and methodological considerations,'' in {\em ACM
  SIGARCH Computer Architecture News}, vol.~23, pp.~24--36, ACM, 1995.

\bibitem{splash2x}
``Splash-2x benchmark suite.''
  \url{http://parsec.cs.princeton.edu/parsec3-doc.htm\#splash2x}.

\bibitem{jimenez2003fast}
D.~A. Jim{\'e}nez, ``Fast path-based neural branch prediction,'' in {\em
  Microarchitecture, 2003. MICRO-36. Proceedings. 36th Annual IEEE/ACM
  International Symposium on}, pp.~243--252, IEEE, 2003.

\bibitem{Andre2006Tage}
A.~Seznec and P.~Michaud, ``A case for (partially)-tagged geometric history
  length predictors,'' {\em Journal of Instruction-Level Parallelism (JILP)},
  vol.~8, 2006.

\bibitem{Andre2014tagescl}
A.~Seznec, ``Tage-sc-l branch predictors.''
  \url{http://www.jilp.org/cbp2014/code/AndreSeznec.tar.gz}, 2014.

\end{thebibliography}

\end{document}